\documentclass{aa}  

\usepackage{graphicx}
\usepackage{txfonts}
\usepackage{natbib}
\newcommand{\um}{\,$\mu$m}
\newcommand{\kms}{km\,s$^{-1}$}
\newcommand{\nii}{[N\,{\sc ii}]}
\newcommand{\cii}{[C\,{\sc ii}]}
\newcommand{\ciiiso}{[$^{13}$C\,{\sc ii}]}
\newcommand{\ci}{[C\,{\sc i}]}
\newcommand{\oi}{[O\,{\sc i}]}
\newcommand{\transl}{$^3$P$_1$-$^3$P$_0$}
\newcommand{\transu}{$^3$P$_2$-$^3$P$_1$}
\newcommand{\hii}{H\,{\sc ii}}
\newcommand{\hh}{H$_2$}
\newcommand{\cc}{cm$^{-3}$}
\newcommand{\coiso}{$^{13}$CO}
\newcommand{\cplus}{C$^+$}
\newcommand{\tex}{$T_\mathrm{ex}$}
\bibpunct{(}{)}{;}{a}{}{,}
\begin{document}

   \title{Velocity resolved \cii, \ci, and CO observations of the N159 star-forming region in the Large Magellanic Cloud: a complex velocity structure and variation of the column densities}

   \titlerunning{Velocity resolved \cii, \ci, and CO observations of the N159 star-forming region}

%   \subtitle{}

   \author{Yoko Okada \inst{1}
          \and
          Miguel Angel Requena-Torres \inst{2}
          \and
          Rolf G\"{u}sten \inst{2}
          \and
          J\"{u}rgen Stutzki \inst{1}
          \and
          Helmut Wiesemeyer \inst{2}
          \and
          Patrick P\"{u}tz \inst{1}
          \and
          Oliver Ricken \inst{2}
          }

   \institute{I. Physikalisches Institut der Universit\"{a}t zu K\"{o}ln, Z\"{u}lpicher Stra{\ss}e 77, 50937 K\"{o}ln, Germany
              \email{okada@ph1.uni-koeln.de}
              \and
              Max-Planck-Institut f\"{u}r Radioastronomie, Auf dem H\"{u}gel 69, 53121 Bonn, Germany
             }

   \date{Received; accepted}

% \abstract{}{}{}{}{} 
% 5 {} token are mandatory
 
  \abstract
   {The \cii\ 158\um\ fine structure line is one of the dominant cooling lines in star-forming active regions. Together with models of photon-dominated regions, the data is used to constrain the physical properties of the emitting regions, such as the density and the radiation field strength.  According to the modeling, the \cii\ 158\um\ line integrated intensity compared to the CO emission is expected to be stronger in lower metallicity environments owing to lower dust shielding of the UV radiation, a trend that is also shown by spectral-unresolved observations.  In the commonly assumed clumpy UV-penetrated cloud scenario, the models  predict a \cii\ line profile similar to that of CO.  However, recent spectral-resolved observations by Herschel/HIFI and SOFIA/GREAT (as well as the observations presented here) show that the velocity resolved line profile of the \cii\ emission is often very different from that of CO lines, indicating a more complex origin of the line emission including the dynamics of the source region.}
   {The Large Magellanic Cloud (LMC) provides an excellent opportunity to study in great detail the physics of the interstellar medium (ISM) in a low-metallicity environment  by spatially resolving individual star-forming regions.  The aim of our study is to investigate the physical properties of the star-forming ISM in the LMC by separating the origin of the emission lines spatially and spectrally.  In this paper, we focus on the spectral characteristics and the origin of the emission lines, and the phases of carbon-bearing species in the N159 star-forming region in the LMC.}
   {We mapped a 4\arcmin$\times$(3\arcmin--4\arcmin) region in N159 in \cii\ 158\um\ and \nii\ 205\um\ with the GREAT instrument on board  SOFIA.  We also observed CO(3-2), (4-3), (6-5), \coiso(3-2), and \ci\ \transl\ and \transu\ with APEX.  All spectra are velocity resolved.}
   {The emission of all transitions observed shows a large variation in the line profiles across the map and in particular between the different species.  At most positions the \cii\ emission line profile is substantially wider than that of CO and \ci.  We estimated the fraction of the \cii\ integrated line emission that cannot be fitted by the CO line profile to be 20\% around the CO cores, and up to 50\% at the area between the cores, indicating a gas component that has a much larger velocity dispersion than the ones probed by the CO and \ci\ emission.  We derived the relative contribution from \cplus, C, and CO to the column density in each velocity bin.  The result clearly shows that the contribution from \cplus\ dominates the velocity range far from the the velocities traced by the dense molecular gas.  Spatially, the region located between the CO cores of N159~W and E has a higher fraction of \cplus\ over the whole velocity range.  We estimate the contribution of the ionized gas to the \cii\ emission using the ratio to the \nii\ emission, and find that the ionized gas contributes $\leq 19$\% to the \cii\ emission at its peak position, and $\leq 15$\% over the whole observed region.  Using the integrated line intensities, we present the spatial distribution of $I_\textrm{[CII]}/I_\textrm{FIR}$.}
   {This study demonstrates that the \cii\ emission in the LMC N159 region shows significantly different velocity profiles from that of CO and \ci\ emissions, emphasizing the importance of velocity resolved observations in order to distinguish different cloud components. }

   \keywords{ISM: lines and bands --
             ISM: kinematics and dynamics --
             Galaxies: Magellanic Clouds --
             ISM: individual objects: N159}

   \maketitle
%
%________________________________________________________________

\section{Introduction}

\begin{table*}
\caption{Observation parameters.\label{table:obsparam}}
\centering
\begin{tabular}{cccccccc}
\hline\hline
Line & Frequency [GHz] & Instrument & $\eta_\mathrm{f}^\mathrm{a}$ & $\eta_\mathrm{mb}^\mathrm{b}$ & HPBW$^\mathrm{c}$ [\arcsec] & $t_\mathrm{eff}^\mathrm{d}$ [s] & $\sigma_\mathrm{rms}^\mathrm{h}$ [K]\\
\hline
\coiso(3-2) & 330.5879653 & FLASH$^+$ & 0.95 & 0.69 & 19.0 & 20--50 & 0.06\\
CO(3-2) & 345.7959899 & FLASH$^+$ & 0.95 & 0.69 & 18.2 & 30--60 & 0.04\\
CO(4-3) & 461.0407682 & FLASH$^+$ & 0.95 & 0.61 & 13.6 & 30--120 & 0.11\\
\ci\ \transl & 492.1606510 & FLASH$^+$ & 0.95 & 0.6 & 12.8 & 100--250 & 0.11\\
CO(6-5) & 691.4730763 & CHAMP$^+$ LFA & 0.95 & 0.42 & 8.8 & 200--700$^\mathrm{e}$, 500--3500$^\mathrm{f}$ & 0.08$^\mathrm{e}$, 0.04$^\mathrm{f}$\\
\ci\ \transu & 809.3419700 & CHAMP$^+$ HFA & 0.95 & 0.38 & 7.7 & 100--500$^\mathrm{e}$, 500-3500$^\mathrm{f}$ & 0.51$^\mathrm{e}$, 0.14$^\mathrm{f}$\\
\nii & 1461.1338000 & GREAT L1 & 0.97 & 0.67 & 18.3 & (50--200)$^\mathrm{g}$ & (0.10)$^\mathrm{g}$ \\
\cii & 1900.5369000 & GREAT L2 & 0.97 & 0.67 & 14.1 & 9--24 & 0.30\\
\hline\\
\end{tabular}
\begin{list}{}{\setlength{\itemsep}{0ex}}
\item[$^\textrm{a}$]Forward efficiency.\ \ \ $^\textrm{b}$ Main beam efficiency.\ \ \ $^\textrm{c}$ Half power beam width.
\item[$^\textrm{d}$]Typical ($> 80$\% of the pixels) effective integration time for one pixel in a 20\arcsec\ resolution map.  For data with CHAMP$^+$, contributions from different pixels are counted as an independent integration time, resulting in a significant decrease of $t_\mathrm{eff}$ at the edge, that is covered by a small number of pixels only.
\item[$^\textrm{e}$]N159~E.\ \ \ $^\textrm{f}$ N159~W.\ \ \ $^\textrm{g}$ In 50\arcsec\ resolution map.\ \ \ $^\textrm{h}$ Median of the baseline noise.
\end{list}
\end{table*}

The \cii\ 158\um\ fine structure transition is one of the dominant cooling lines in photon-dominated regions \citep[PDRs;][]{TH85I,Sternberg1995}, and it is used to constrain their physical properties.  Because of its strength, it also has  a great potential for tracing star formation activity in external galaxies and for estimating the star formation rate \citep[][and references therein]{deLooze2011}.  On the other hand, changes in the metallicity affect the abundance of the major coolants including carbon, the electron densities in PDRs, the dust abundance that determines the UV penetration, the photoelectric heating rate, and the \hh\ formation rate, resulting in a larger \cplus\ layer in the  low-metallicity environment \citep{Bolatto1999,Roellig2006}.  This metallicity dependence is also shown observationally as a two-order-of-magnitude difference in the intensity ratio of \cii\ and CO among galaxies with different metallicities \citep{Mochizuki1994,Poglitsch1995,Israel1996,Madden1997, Madden2011}.  In order to understand the extragalactic \cii\ emission, which is the superimposed emission from different physical environments, our interpretation needs to bridge detailed Galactic studies to observations of marginally or non-spatially resolved external galaxies. The Large Magellanic Cloud (LMC) is one of the most important targets  because its different metallicity from our Galaxy \citep[$\sim 1/3$ of solar;][]{Carrera2008} enables us to investigate the influence of the metallicity on the thermal balance and chemical composition in the ISM, and its proximity (50~kpc as a median distance of 729 references; NASA/IPAAC Extragalactic Database) allows spatially resolved studies of the large star-forming regions visible in the LMC.

In most cases, integrated line intensities of \cii\ and CO emission are used as diagnostics in external galaxies.  However,  velocity resolved  mapping observations of the \cii\ emission in Galactic regions have recently been made with the Heterodyne Instrument for the Far-Infrared (HIFI) on {\it Herschel} and are being made with the German REceiver for Astronomy at Terahertz Frequencies (GREAT) on board the Stratospheric Observatory for Infrared Astronomy (SOFIA), and they show complex line profiles of the \cii\ emission, typically substantially wider than expected from the \ci\ and low- or mid-J CO emission profiles. The velocity information is thus essential in order to identify the origin of the \cii\ emission \citep{PerezBeaupuits2012,Carlhoff2013}, and the significant difference in the velocity profile of the \cii\ emission from the low- or mid-J CO emission indicates that a substantial fraction of the \cii\ emission is accelerated relative to the quiescent material, e.g.,  it undergoes ablation or is blown off \citep{Dedes2010,Mookerjea2012,Okada2012,Schneider2012,Simon2012,Pilleri2012b}.  These results emphasize the need of velocity resolved observations to distinguish different velocity components with their different physical conditions.

Before the HIFI and GREAT observations only the pioneering heterodyne observations from the Kuiper Airborne Observatory (KAO) were available.  \citet{Boreiko1991} observed 17 locations in the LMC and showed that the \cii\ emission is approximately 50\% wider than the CO(2-1) line.  However, investigating its spatial variation and the line profile in detail requires mapping observations and a higher signal-to-noise ratio (S/N) than was available at that time;  HIFI provided the necessary higher S/N, but the mapping ability was still limited and allowed   only a few strip maps to be observed across a few prominent regions in the LMC.

Since CO lines are observable from the ground, more velocity resolved observations  have been done for them.  In the N159 star-forming region, which we focus on in this paper, physical properties of the clouds have been studied using low-J CO mapping observations \citep{Johansson1998,Minamidani2008,Mizuno2010,Minamidani2011} or using single point observations of CO(4-3), CO(7-6), and  \ci\ \transl\ and \transu\ emissions \citep{Pineda2008}, but detailed line profiles have not been discussed.  \citet{Bolatto2000} present a large-scale map and velocity resolved spectra of \ci\ \transl, but with a spatial resolution of only 4\arcmin, which is insufficient to resolve the individual source components spatially.

The aim of our GREAT core project on the LMC and Small Magellanic Cloud (SMC) is to study physical properties by distinguishing the origin of the emission lines using velocity profiles in active regions in the LMC and  SMC and to investigate the effect of the low-metallicity environment on the properties and evolution of the interstellar medium (ISM).  In this paper, we present observations in N159, focusing on the velocity profiles and the origin of the emission lines, and present the column density ratio between \cplus, C, and CO using simple assumptions.  We do not deal  with any elaborate PDR modeling or with the metallicity effects.  A full PDR model analysis will be presented in a follow-up paper (Okada et al. in prep.).  N159 is located south of 30 Dor as a part of the molecular ridge \citep{Cohen1988,Mizuno2001}.  Three CO cores are identified \citep[N159~W, E, and S;][]{Johansson1998}.  N159~S shows no evidence of star-forming activity, while W and E possess O-type stars and \hii\ regions, where massive stars formed a few Myr ago \citep{Chen2010}.  The results of 30 Dor and N66 in the SMC are presented in a separate paper (Requena et al. in prep.).

%__________________________________________________________________

\section{Observations and data reduction}
\subsection{\cii\ 158\um\ and \nii\ 205\um\ observations with SOFIA/GREAT}

We  mapped 4\arcmin$\times$(3\arcmin--4\arcmin) in the N159 star-forming region in the LMC, observing the \cii\ emission at 1900.5369~GHz (158\um) using the German REceiver for Astronomy at Terahertz Frequencies \citep[GREAT\footnote{GREAT is a development by the MPI f\"{u}r Radioastronomie and the KOSMA / Universit\"{a}t zu K\"{o}ln, in cooperation with the MPI f\"{u}r Sonnensystemforschung and the DLR Institut f\"{u}r Planetenforschung};][]{Heyminck2012} on board the Stratospheric Observatory for Infrared Astronomy \citep[SOFIA;][]{Young2012} in July 2013, as part of the open time and guaranteed time in cycle 1 observations.  Observations were performed during four flights of SOFIA's first southern deployment to New Zealand and the L2 channel was tuned to \cii.  During the first two flights, we observed in parallel \nii\ 205\um\ with the L1 channel.  The observations were made in total-power on-the-fly (OTF) mode, with 0.5 to 2 second integration time per dump, at 6\arcsec\ step size.  The total flight time for N159 observations was about 4 hours and the total integration time on source was about 1 hour.  The OFF position for the observations was located southwest  of the observed region (05$^\mathrm{h}$38$^\mathrm{m}$53$^\mathrm{s}$.3,$-69^\circ$46\arcmin29\arcsec\ (J2000)), which is the same OFF position used in \citet{Pineda2008}.

Pointing on SOFIA is done via a combination of fast inertial gyro stabilization of the telescope and regular corrections on the optical guide cameras. Experience shows that pointing and tracking is better than 2 arcsecs. The alignment between the telescope optical axis, i.e., the guide cameras, and the boresight of the GREAT instrument is checked and verified at the beginning of each observing campaign after installation of the GREAT instrument by observing a suitable planet, and was 1.5 arcsecs.

The data from the extended bandwidth Fourier transform spectrometer (XFFTS; 2.5~GHz bandwidth with 44~kHz resolution) were calibrated by the standard GREAT pipeline \citep{Guan2012}.  Details of the observations are listed in Table~\ref{table:obsparam}.  A polynomial of second order was used to fit the baseline.  We checked the results obtained with polynomials of first to fourth order as a baseline, and found that a first-order polynomial results in higher baseline noise, fourth-order polynomials introduce artifacts, and second- and third-order polynomials make no significant difference.  After baseline subtraction, the data were spectrally resampled to 1~\kms\ resolution and spatially resampled to 20\arcsec\ resolution using the baseline noise as the weight.  In the analysis of the very weak \nii\ emission, we resampled the data to 2~\kms\ and 50\arcsec\ resolution to obtain a better S/N.  Data reduction up to the spectral resampling was done by the CLASS software \citep{Pety2005}, and the spatial resampling and further analysis was done with IDL.

\subsection{CO and \ci\ observations with APEX}

\begin{figure*}
\centering
\includegraphics[bb=40 10 424 350,width=0.560\hsize,clip]{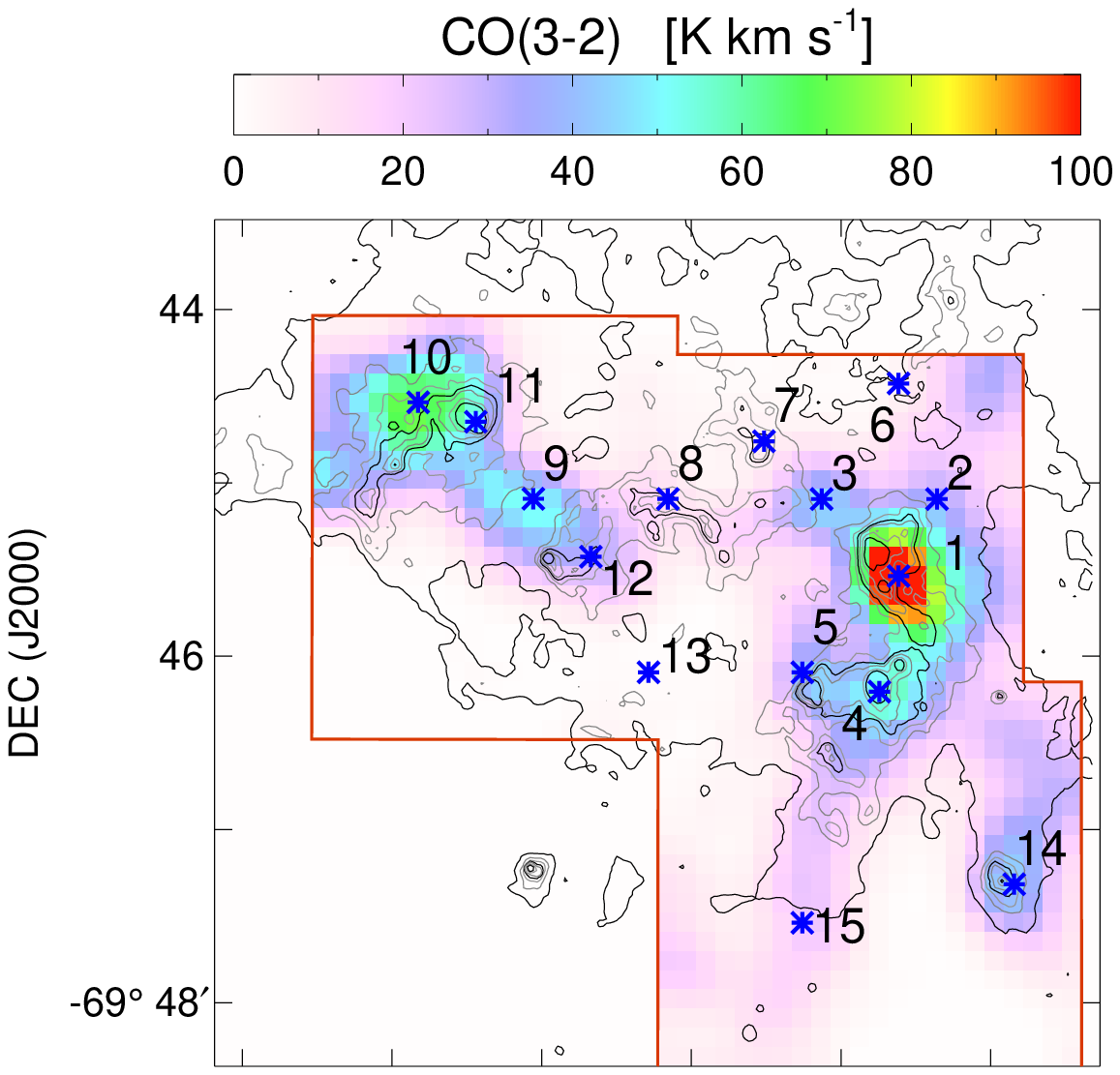}
\hspace*{-1cm}
\includegraphics[bb=95 10 424 350,width=0.480\hsize,clip]{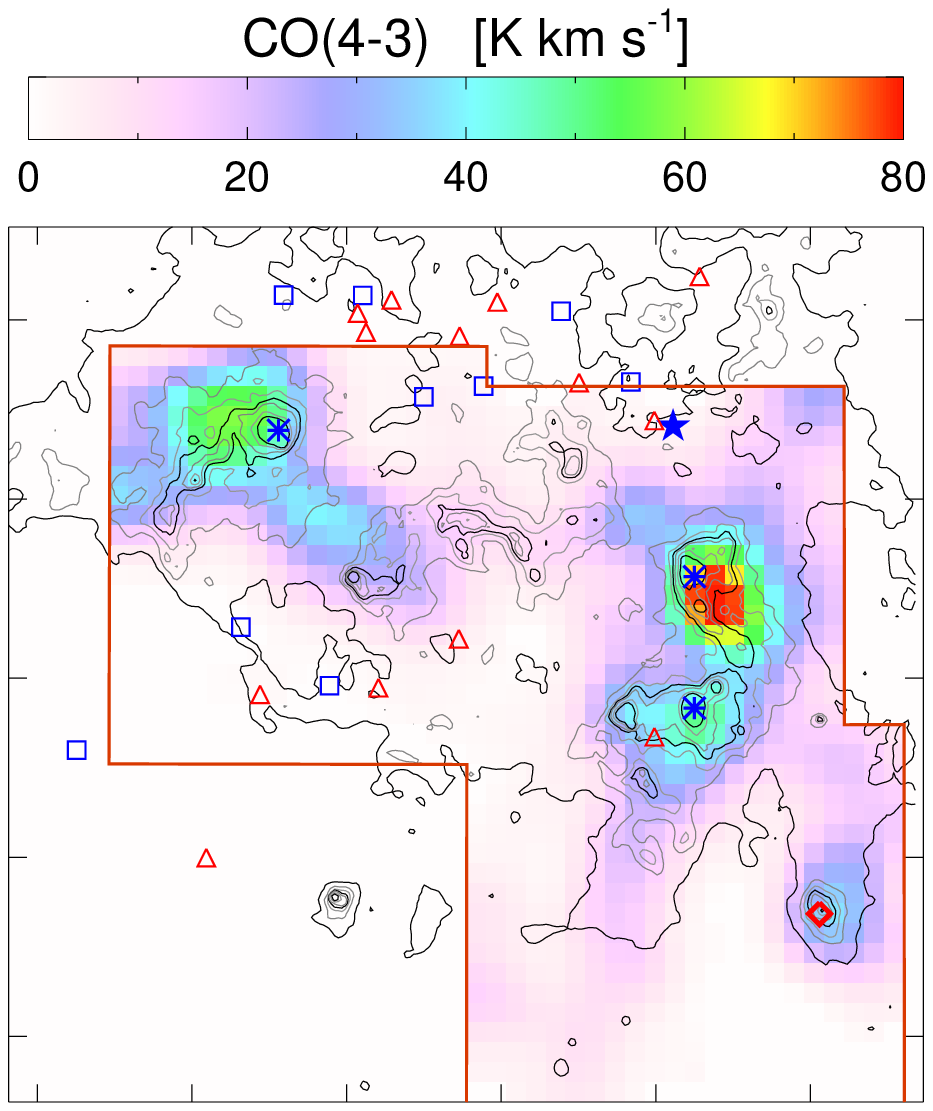}
\includegraphics[bb=40 0 424 360,width=0.560\hsize,clip]{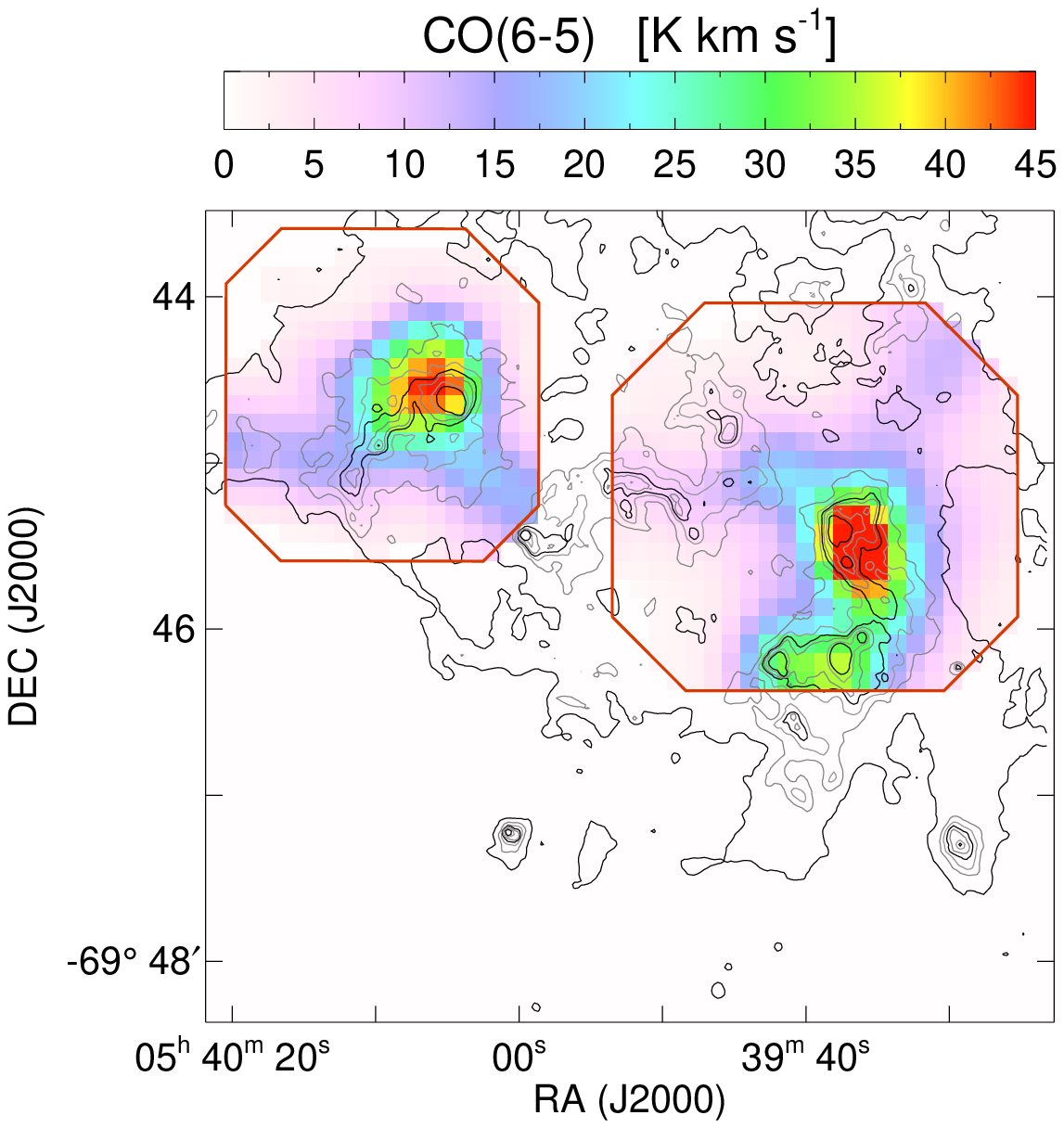}
\hspace*{-1cm}
\includegraphics[bb=95 0 424 360,width=0.480\hsize,clip]{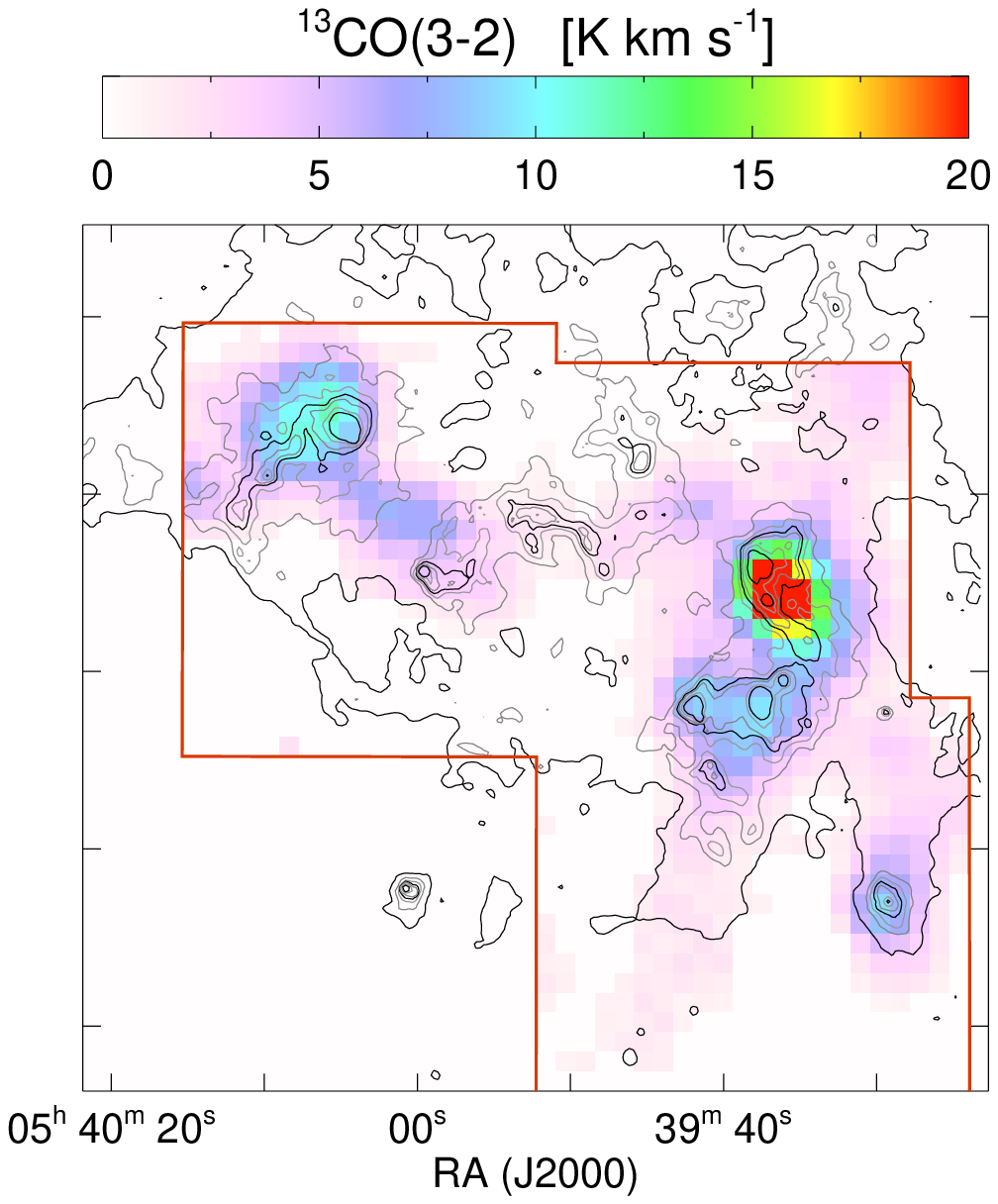}
\caption{Integrated ($220$ to $250$~\kms) intensity maps (colors, 20\arcsec\ resolution) overlayed with the contours of IRAC 8\um. The last (eighth) panel shows a three-color image composed of CO(3-2) (blue), \ci\ \transl\ (green), and \cii\ (red).  The red lines outline the observed area.  Blue asterisks in the CO(3-2) and \cii\ maps mark the positions whose spectra are shown in Fig.~\ref{fig:spectra}.  Open blue squares and red triangles in the CO(4-3) map show the position of stars with a spectral type of O7 or earlier and O7 to B1, respectively, from \citet{Farina2009}, the filled blue  star marks the position of LMC X-1, the blue asterisks are compact \hii\ regions \citep{Indebetouw2004}, and the red diamond marks the position of an H$_2$O maser \citep{Lazendic2002}.\label{fig:integ_map}}
\end{figure*}

\addtocounter{figure}{-1}

\begin{figure*}
\centering
\includegraphics[bb=40 5 424 355,width=0.560\hsize,clip]{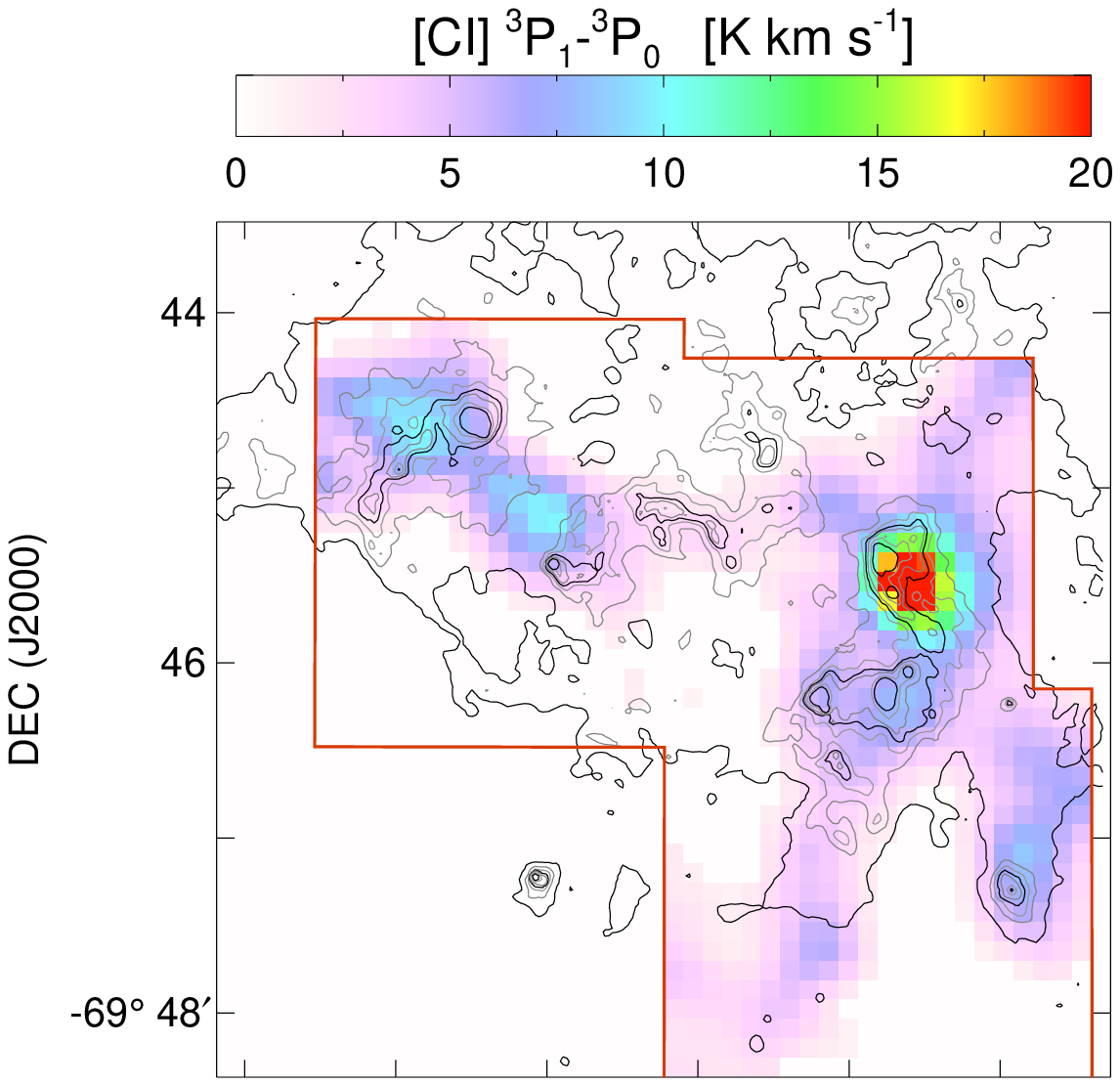}
\hspace*{-1cm}
\includegraphics[bb=95 5 424 355,width=0.480\hsize,clip]{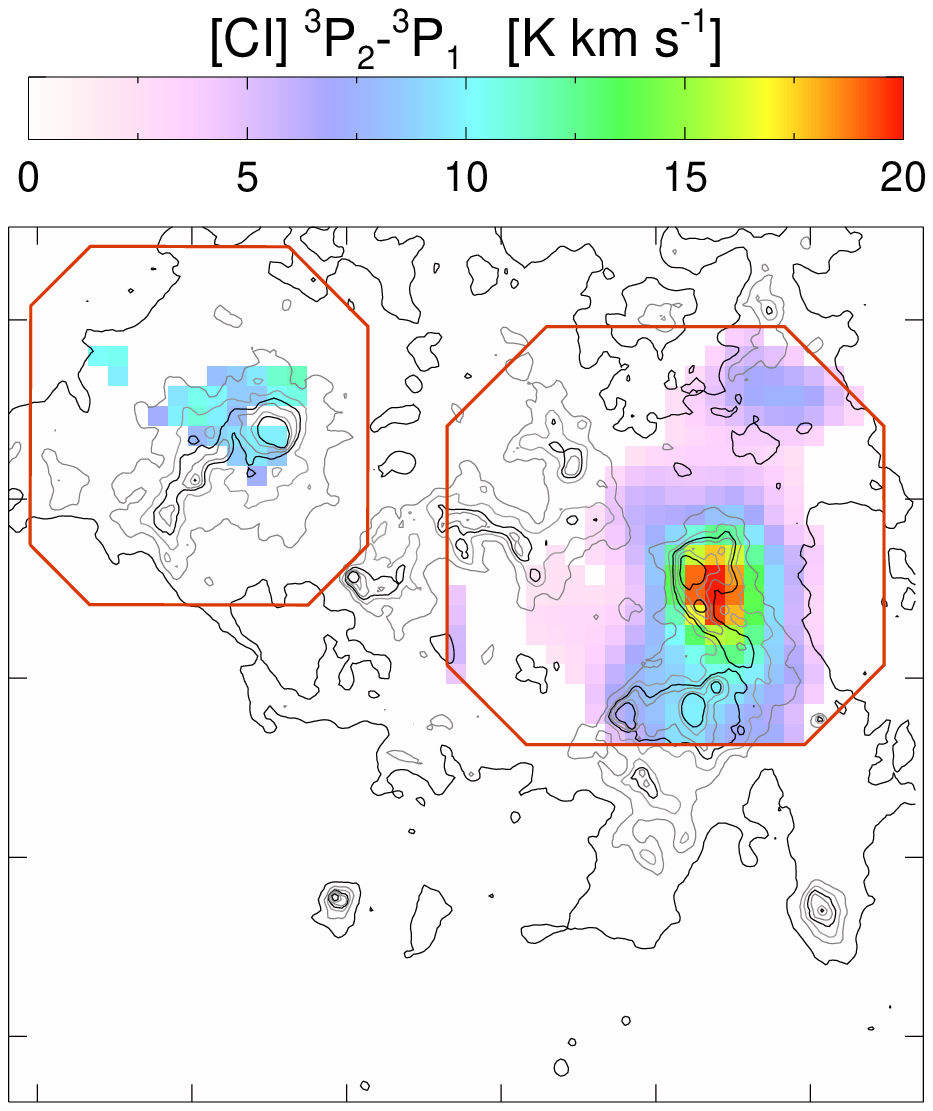}
\includegraphics[bb=40 0 424 350,width=0.560\hsize,clip]{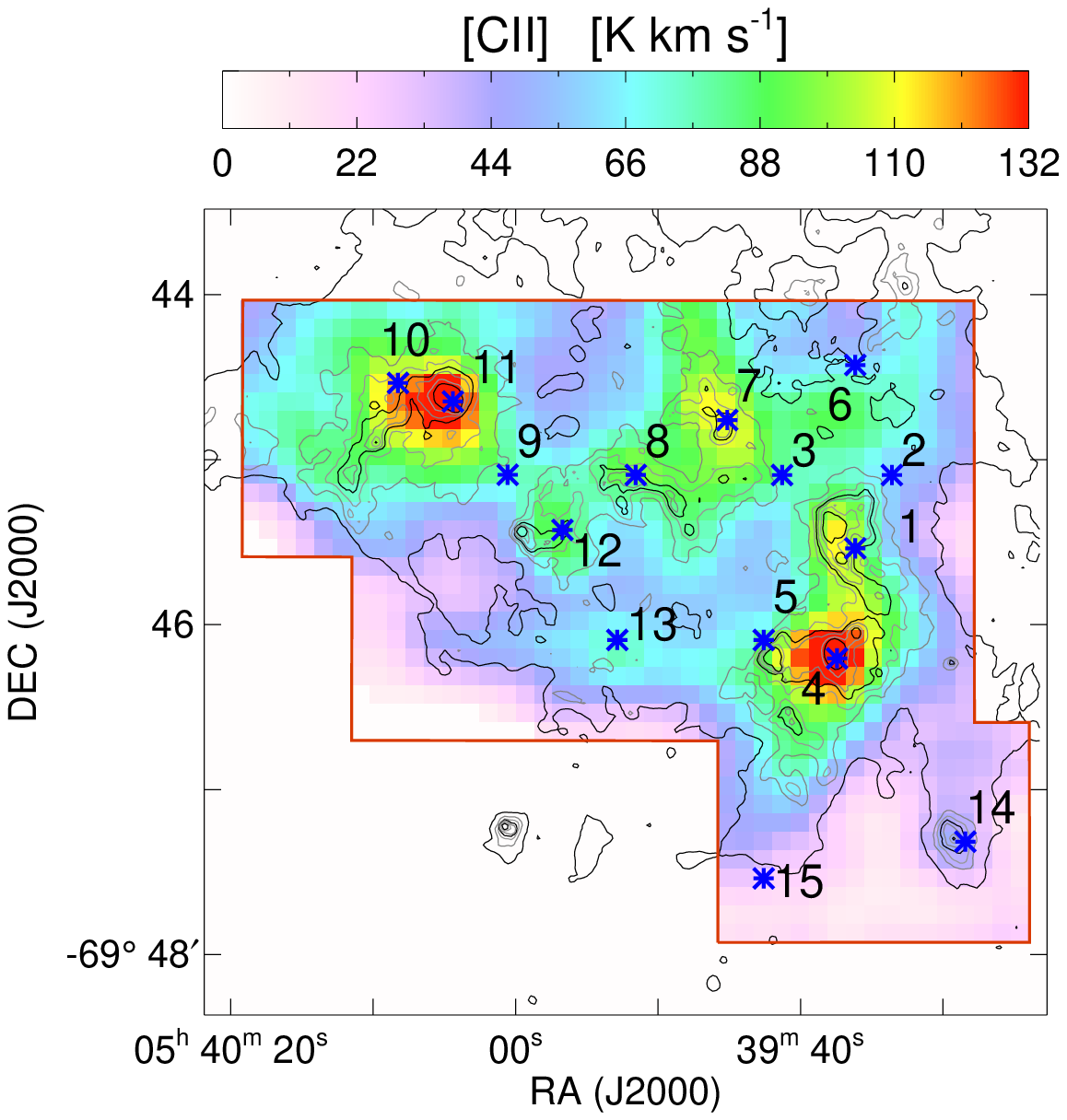}
\hspace*{-1cm}
\includegraphics[bb=95 0 424 350,width=0.480\hsize,clip]{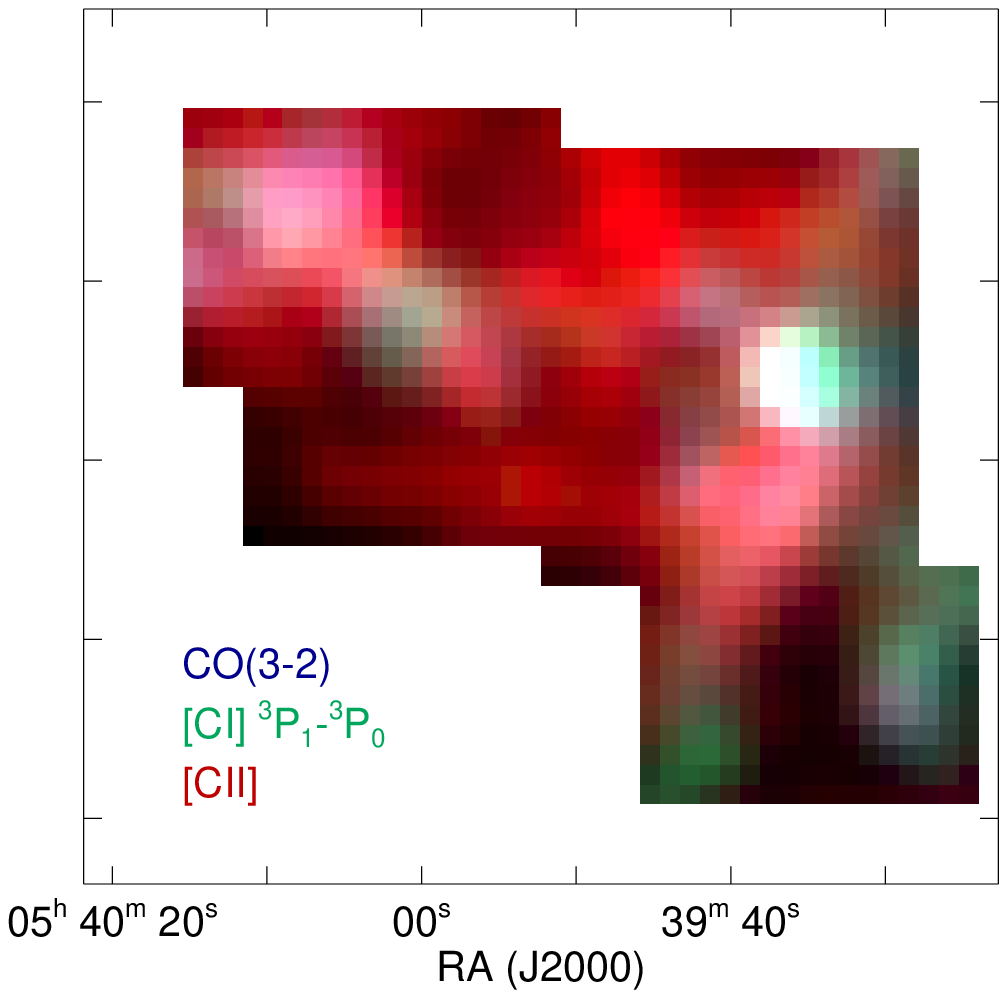}
\caption{\it{(continued)}}
\end{figure*}

In addition to the GREAT \cii\ and \nii\ mapping, we carried out complementary observations with the Atacama Pathfinder Experiment \citep[APEX\footnote{APEX is a collaboration between the Max-Planck-Institut fur Radioastronomie, the European Southern Observatory, and the Onsala Space Observatory.};][]{Guesten2006} of CO(3-2), CO(4-3), \coiso(3-2), and \ci\ \transl\ using the MPIfR heterodyne receiver FLASH$^+$ \citep{Klein2014}, and CO(6-5) and \ci\ \transu\ using the  Carbon Heterodyne Array of the MPIfR \citep[CHAMP$^+$;][]{Kasemann2006} in June, July, September, and November 2012.  The backends of FLASH$^+$ are FFTSs with the spectral resolution of 38~kHz and 76~kHz for the 345~GHz and 460~GHz band, respectively.  CHAMP$^+$ has two hexagonal arrays each with 7 pixels, simultaneously operated for the  620--720~GHz and the 780--950~GHz frequency range, and the backends are XFFTSs (1.5~GHz bandwidth with 212~kHz resolution).  All observations were made in the OTF mode, and the OFF position is the same as in the GREAT observations.  Observational parameters are listed in Table~\ref{table:obsparam}.  Pointing was established on nearby R-Dor with CO lines, repeated every hour, and the pointing uncertainty is less than 2.5\arcsec.

\citet{Johansson1998} mapped $1\arcmin\times 1\arcmin$ around the brightest core of N159~W in CO(3-2) with the Swedish-ESO Submillimetre Telescope (SEST), but it does not cover the whole region of our \cii\ observations.  \citet{Minamidani2008} show a larger CO(3-2) map including another prominent core of N159~E obtained with the Atacama Submillimeter Telescope Experiment (ASTE), but these are position switched observations with a grid of 10\arcsec\ to 40\arcsec\ and are not fully sampled.  Therefore, we included CO(3-2) in our APEX observations in order to have sufficient complementary data.

A linear baseline was fitted in the ranges 150 - 225~\kms \ and 255 - 350~\kms\ for FLASH$^+$, and 100 - 220~\kms \ and  255 - 400~\kms\ for CHAMP$^+$.  After discarding spectra with excess noise, the data were spectrally resampled to 1~\kms\ and spatially to 20\arcsec, identical to the resolution of the \cii\ data.  As with the GREAT data, reduction up to the spectral resampling was performed using the CLASS software, and the spatial resampling and further analysis was done with  IDL.

%______________________________________________________________

\section{Results and discussion}
\subsection{Spatial and velocity structures}\label{subsec:result_structure}

\begin{figure*}
\centering
\includegraphics{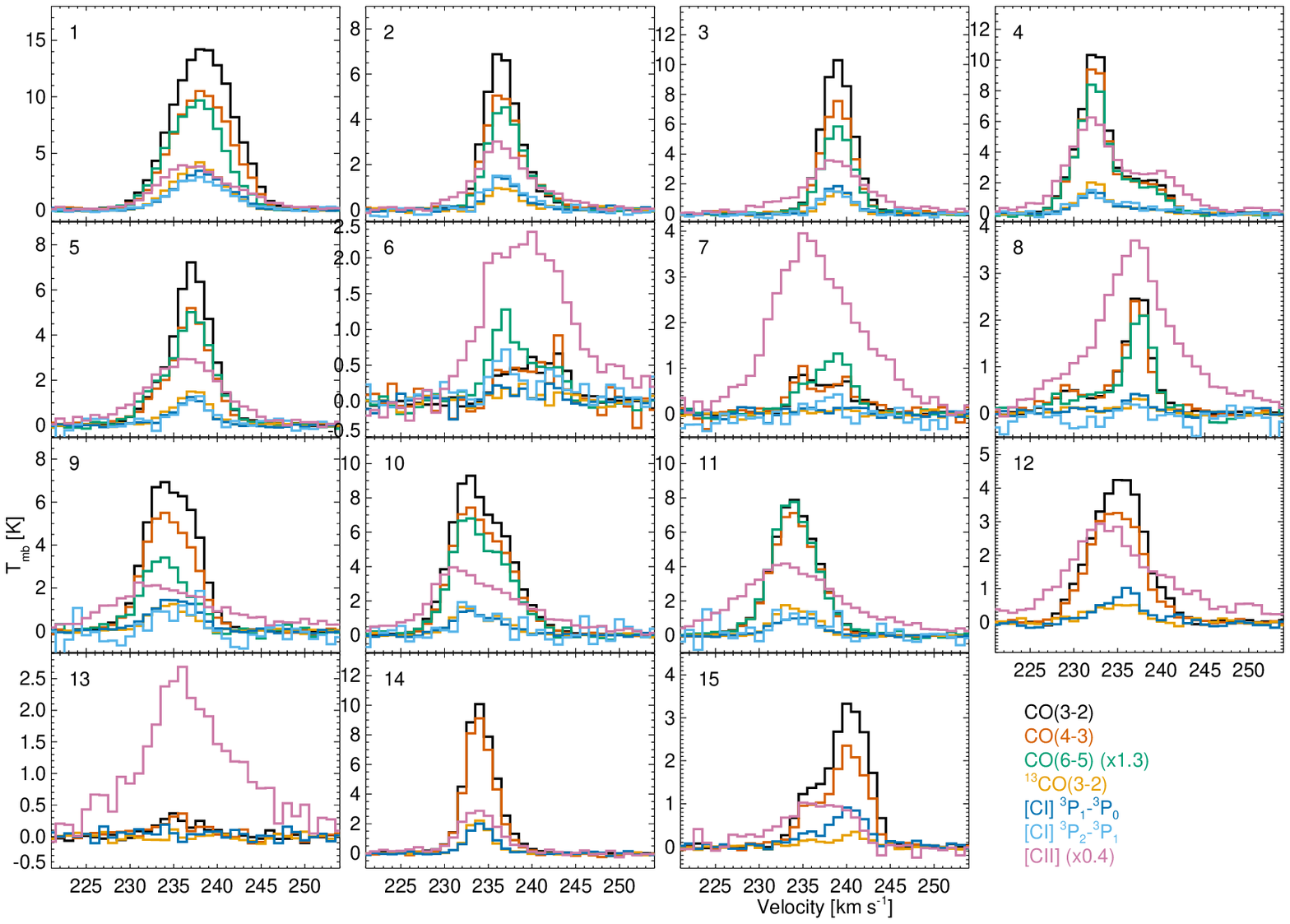}
\caption{Spectra at selected positions marked in the CO(3-2) panel in Fig.~\ref{fig:integ_map}.
\label{fig:spectra}}
\end{figure*}

\begin{figure*}
\centering
\includegraphics[bb=0 0 504 340,width=\hsize,clip]{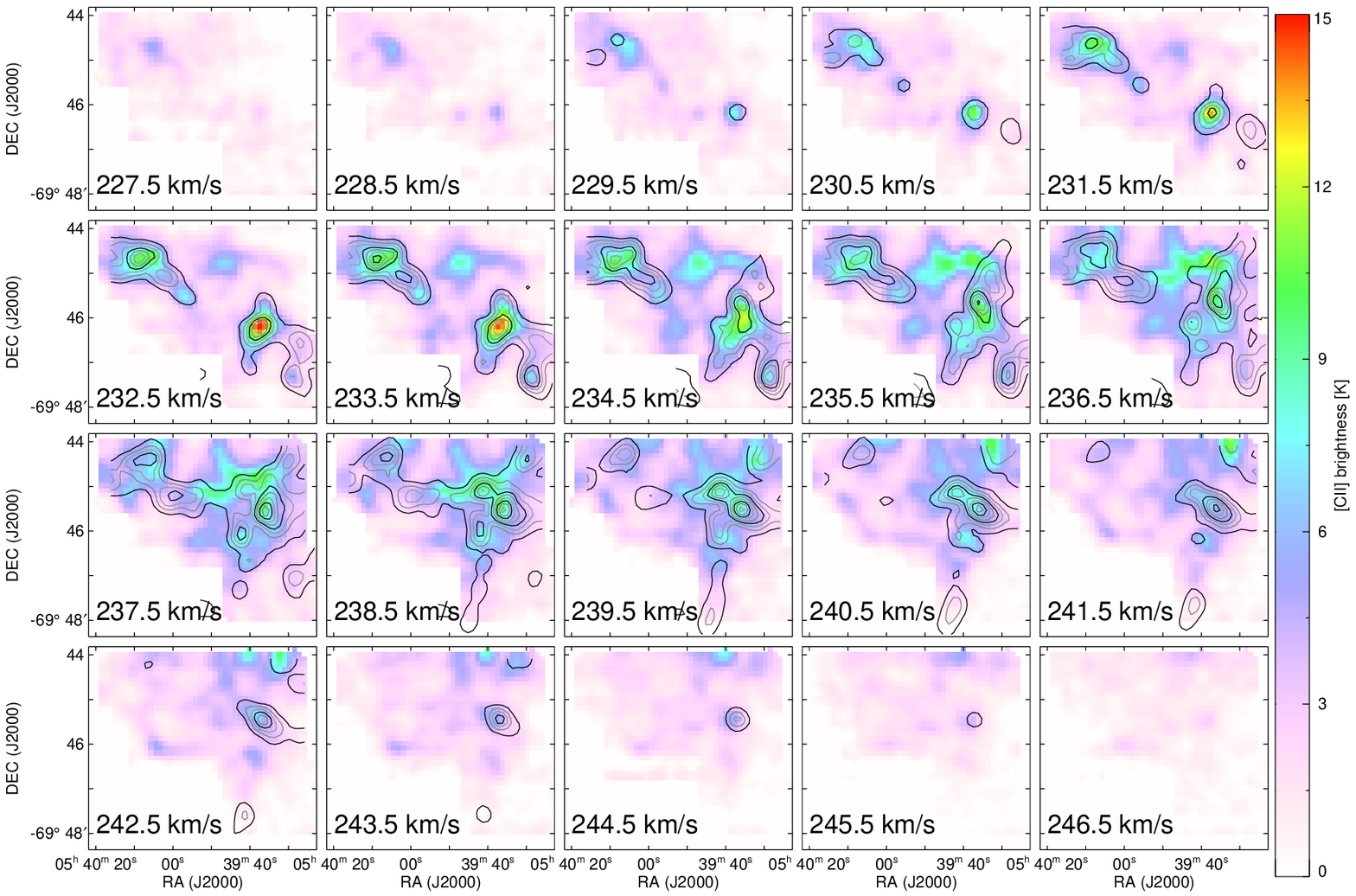}
\caption{1~\kms\ wide channel maps of the \cii\ emission overlayed with the contours of the CO(3-2) emission.  The contours start from 1.5~K with a spacing of 1.5~K.  The central velocity is written in each panel.\label{fig:channelmap}}
\end{figure*}

\begin{figure*}
\centering
\includegraphics[bb=0 0 420 350,width=0.8\hsize,clip]{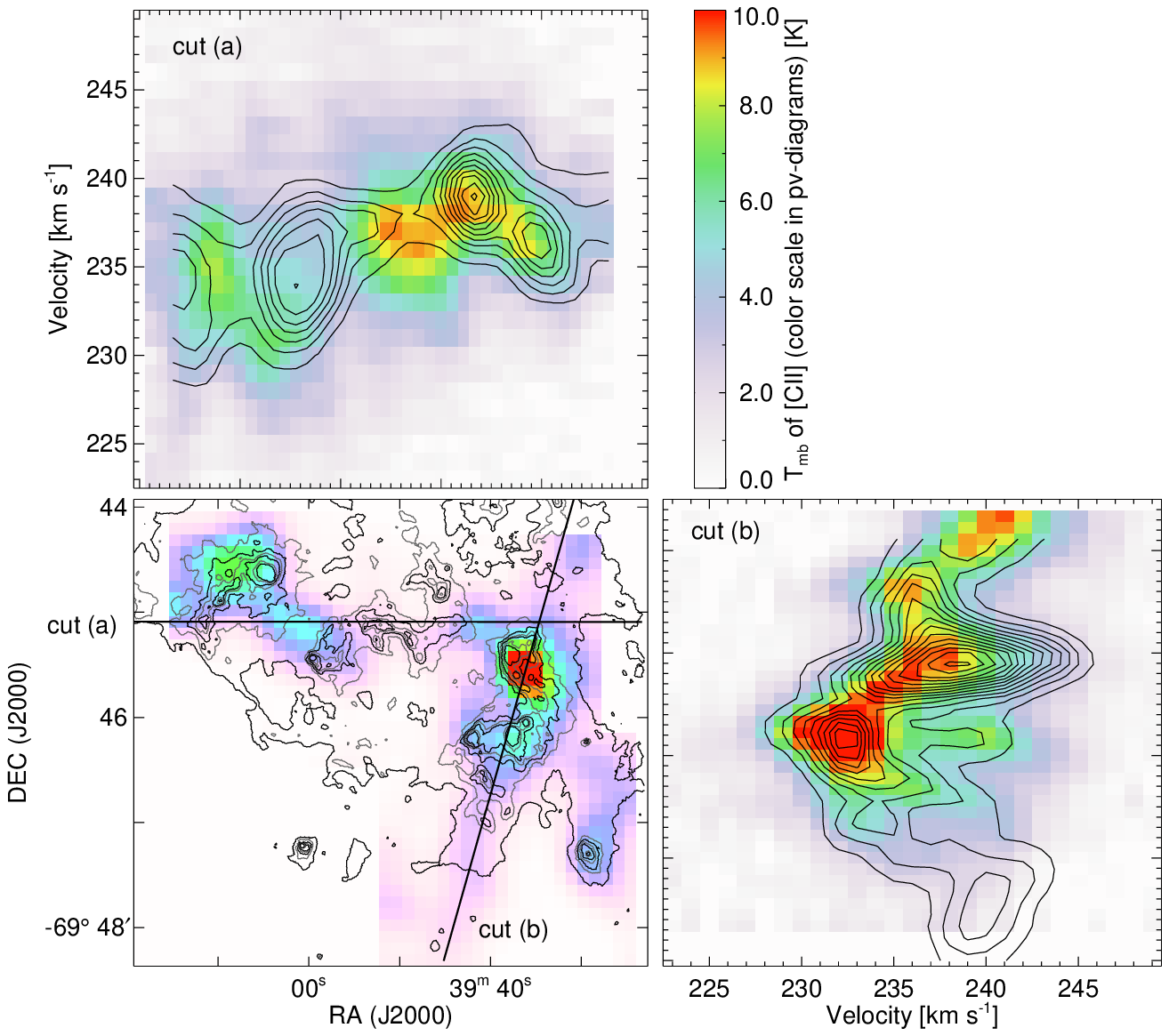}
\caption{(lower left) CO(3-2) integrated intensity map overlayed with the contours of IRAC 8\um\ (same as Fig.~\ref{fig:integ_map}), together with two lines marking the two cuts, along which the p-v diagrams are shown in the upper left and lower right panel. (upper left / lower right) The p-v diagrams of the \cii\ (color) and CO(3-2) (contour) in $T_\mathrm{mb}$ [K].  The contour spacing is 1~K.  The position axis is horizontal in the upper left panel (cut a) and vertical in the lower right panel (cut b), so that the position aligns with the position of the lower left panel.\label{fig:pvdiagram}}
\end{figure*}

Figure~\ref{fig:integ_map} shows the integrated intensity maps of CO(3-2), (4-3), (6-5), \coiso(3-2), \ci\ \transl\ and \transu, and \cii, overlayed with the contours of IRAC 8\um\ from the {\it Spitzer} data archive.  We estimate the uncertainty of the integrated intensity ($\sigma_I$ [K~\kms]) based on the baseline noise of the spectrum at each position of the resampled map ($\sigma_\mathrm{rms}$, Table~\ref{table:obsparam}), and the velocity range used for the integrated intensity (220--250~\kms).  In Fig.~\ref{fig:integ_map}, pixels that have an integrated intensity smaller than $3\sigma_I$ were masked.  

As shown in the following, the velocity structure strongly varies across the observed area, and from line to line.  Therefore, the integrated intensity map has to be combined with spectral information to interpret the spatial structure.  Spectra at selected positions (marked in the CO(3-2) and \cii\ maps of Fig.~\ref{fig:integ_map}) are shown in Fig.~\ref{fig:spectra}; Fig.~\ref{fig:channelmap} shows the channel maps of the \cii\ and CO(3-2) emission; and the position-velocity (p-v) diagram of CO(3-2) and \cii\ along two selected cuts are shown in Fig.~\ref{fig:pvdiagram}.

In a first step of our analysis we confirmed that our CO(3-2) map reproduces bright cores shown by \citet{Johansson1998} and \citet{Minamidani2008}.  At N159~W (position 1 in Figs.~\ref{fig:integ_map} and \ref{fig:spectra}, p1 in the following), CO(3-2) and all other lines show a wider ($\sim 8$~\kms\ in FWHM) velocity profile compared to surrounding regions.  Regions around p1 (p2, p3, and p4) show narrower profiles with different central velocities, which may blend at p1.  The integrated intensity peak of CO(3-2) is shifted southwest compared to the peak in IRAC 8\um, and the IRAC 8\um\ has a cometary shape facing northeast.  The spatial distribution indicates a stratification from the northeast as an ionizing front to the molecular cloud toward the southwest. On the other hand, CO(3-2) in the northeast of N159~W does not completely vanish, but shows a weak peak (p3);  a very weak east-west bridge along IRAC 8\um\ is still seen in CO(3-2) (including p8).  The cut (a) in the p-v diagram of CO(3-2) (Fig.~\ref{fig:pvdiagram}) goes through p2, p3, p8, and p9, and it shows that the weak bridge connects clumps at p3 and p9 in the velocity as well, indicating that these clumps are not independent but rather connected with each other.  The core south of N159~W (p4) is not strong in CO(3-2).  From this core, a long ridge continues toward the south up to the edge of the map with a large velocity gradient (cut (b) in Fig.~\ref{fig:pvdiagram} and p15).  On the western side, there is another north-south ridge connected to N159~W.  It contains a strong core in both CO(3-2) and IRAC 8\um\ at the southern edge (p14, the H$_2$O maser source).  At N159~E, the CO(3-2) peak (p10) shifts to the east compared to IRAC 8\um, and the extension to the southeast is also the east side of the IRAC 8\um\ ridge.  The structure to the southwest from N159~E (p9) fills the gap in IRAC 8\um, and is connected to the bridge coming from N159~W.

The spatial structure of prominent cores in the integrated intensity maps of CO(3-2), CO(4-3), CO(6-5), and \coiso(3-2) is similar, but the relative strength between cores looks different.  For example, the relative strength of p4 to p1 is larger in CO(6-5) than in CO(3-2), indicating a difference in the physical conditions in the different cores.  Moreover, the spectral profile of CO(6-5) shows a clear difference with CO(3-2) at some positions because the different velocity components trace different clumps with different physical properties.  For example at p7 (Fig.~\ref{fig:spectra}), the CO(4-3) and (3-2) emissions have two significant velocity components at $\sim 235$~\kms\ and $\sim 240$~\kms, while the CO(6-5) emission at $\sim 235$~\kms\ just shows a wing, and its $\sim 240$~\kms\ component is very strong in terms of the ratio to CO(4-3) or (3-2).  At p6, both the CO(6-5) and \ci\ \transu\ emission have a peak at $\sim 237$~\kms\ and especially the CO(6-5) emission is strong, while CO(4-3) and (3-2) have a broad velocity profile peaking at $>240$~\kms, and \ci\ \transl\ is even not detected.  Position 6 is close to LMC X-1, which is the X-ray binary including a black hole with a mass of $\sim 11M_\odot$ \citep{Orosz2009} and an O7 III companion \citep{Cowley1995}.  The $\sim 237$~\kms\ component may originate in a clump that  is affected by strong X-ray emission from LMC X-1.

The \ci\ \transl\ map and spectra show an overall similarity with CO(3-2), but the two north-south ridges south of N159~W are stronger in \ci\ \transl.  Position 15 is the peak in \ci\ \transl, whereas the peak there is not clearly seen in other emission lines.  The peak of the western ridge shifts slightly to the north in \ci\ \transl.  The \ci\ \transu\ emission is detected in a wide area of N159~W and around the peak of \ci\ \transl\ in N159~E.  It is detected in only a small area around N159~E, which may be because of the shorter observing time and thus the lower sensitivity there (Table~\ref{table:obsparam}).

The \cii\ emission shows significantly different velocity profiles and integrated intensity maps from the other tracers.  It  typically has a wider velocity profile, except for a few positions.  At p1 and p14, all the lines have a wider and narrower velocity profile than the other positions, respectively, and there is no significant difference in the width between \cii\ and other emissions, except that the peak velocity of the \cii\ emission at p1 is shifted to a smaller velocity (Figs.~\ref{fig:spectra} and \ref{fig:pvdiagram}).  At p9--p12, the \cii\ emission is not only wider than in the other tracers, but  the peak velocity is also shifted by a few \kms.  The integrated intensity of \cii\ spatially matches   the IRAC 8\um: in N159~W, their peaks match around p1 while those of CO and \ci\ emission are shifted southwest, and in N159~E their peaks are instead at p11 in contrast to the CO and \ci\ peak at p10.  An exception is p13, where the \cii\ emission indicates a clump but  there is no corresponding IRAC 8\um\ emission.  All CO and \ci\ emission lines are also very weak, and the H$\alpha$ map by \citet{Chen2010} and the 843~MHz radio continuum map by \citet{Mills1984} show the presence of ionized gas position, so that it is likely that the \cii\ emission at p13 also originates in the ionized gas component. Unfortunately, our \nii\ map, which would allow  the \cii\ contribution from ionized gas to be estimated, does not cover this position.  The ridge-like structure between N159~W and E (including p8) is significant in \cii.  The p-v diagram of cut (a) in Fig.~\ref{fig:pvdiagram} shows that the velocity gradient and the central velocity are similar between CO(3-2) and \cii, indicating that the emission along the cut in the two tracers does not come from fully independent clouds but is related, although the \cii\ emission has a much wider line width.  At p7, the velocity component at $\sim 240$~\kms, which is significant in CO(6-5), is not detectable in the \cii\ emission.

We compare our \cii\ observations with previous studies and archival data from Herschel/PACS.  \citet{Israel1996} mapped the N159 region with FIFI/KAO; they quote a beam solid angle as a 68\arcsec disk, which corresponds to a Gaussian FWHM of 64\arcsec.  We convolved our map to 64\arcsec\ resolution and derive a line integrated intensity of $(6.0$--$6.3)\times 10^{-4}$~erg~cm$^{-2}$~s$^{-1}$~sr$^{-1}$ at the peak in N159~E and W, whereas their result was $(3.5$--$4.0)\times 10^{-4}$~erg~cm$^{-2}$~s$^{-1}$~sr$^{-1}$.  \citet{Boreiko1991} detected the \cii\ emission at N159~FIR(NE) with their far-infrared heterodyne Schottky-mixer receiver on the KAO, showing a peak antenna temperature of 5.3~K and an integrated intensity of $4.8\times 10^{-4}$~erg~cm$^{-2}$~s$^{-1}$~sr$^{-1}$ in the 43\arcsec\ FWHM beam.  Convolving our map to 43\arcsec\ and extracting the spectrum at their measured position (which is 25\arcsec\ off from our peak of N159~E), we obtain a peak main beam temperature of $5.9$~K and an integrated intensity of $5.9\times 10^{-4}$~erg~cm$^{-2}$~s$^{-1}$~sr$^{-1}$.  The value measured with GREAT is 10--20\% higher, but considering that we are comparing the antenna temperature with the main beam temperature for an extended source, the two observations are compatible.  We also retrieved \cii\ observations with PACS from the Herschel data archive, and averaged the spectra in a 20\arcsec\ diameter circle based on the geometrical overlap of the PACS spaxels.  The values derived for N159~W (p1 and p4) and E (p10 and p11) are ($6.1$, $7.7$, $7.4$, and $8.2)\times 10^{-4}$~erg~cm$^{-2}$~s$^{-1}$~sr$^{-1}$ for PACS and ($7.2$, $10.8$, $8.4$, and $10.3)\times 10^{-4}$~erg~cm$^{-2}$~s$^{-1}$~sr$^{-1}$ for GREAT, showing that the PACS intensity is $0.7$--$0.9$ of the GREAT intensity.  We consider this difference to be within the calibration uncertainties of both  instruments.

\subsection{Quantitative analysis of the velocity profiles}
 \label{subsec:gaussianprofile}

To investigate quantitatively the velocity profiles and their differences in different emission lines, we fit the CO(3-2) spectra with two Gaussians, and apply the fitted center velocity and width in the fit to the other emission lines, fitting only their amplitudes. Across the map, there is no position that has more than two clearly separate velocity components.  In order to handle positions where a de-composition into two velocity components is not possible, we first fit the CO(3-2) spectra with two Gaussians with center, width, and amplitude as free parameters, and when the separation of the fitted center of the two Gaussians is less than 1~\kms, or the fitted amplitude of one of the Gaussians is less than the noise of the baseline, we re-fit the spectra with a single Gaussian.  For the positions where the two velocity components are not clearly separated, the fitting may not be unique, but the sum of the integrated intensity of the two components is robust.  Once the width(s) and center(s) of the Gaussian(s) from the CO(3-2) fit are determined, we use these values as fixed parameters in the fitting of the other emission lines, thereby leaving their amplitude as the only free parameter.  In these fits, we restricted the spectral range to be fitted to half of the FWHM of the CO 3-2 fit, as long as there is a data point in this range (otherwise we expand the fitting range to the FWHM) to make sure that the peak of the line is well fitted when the profile of the line is different from that of CO(3-2).  This procedure allows us to identify how well the other emission lines match the velocity structure visible in CO(3-2) and to identify whether the integrated line emission from the other tracers shows evidence for additional velocity components not visible in CO(3-2).

\begin{figure}
\centering
\includegraphics[width=\hsize]{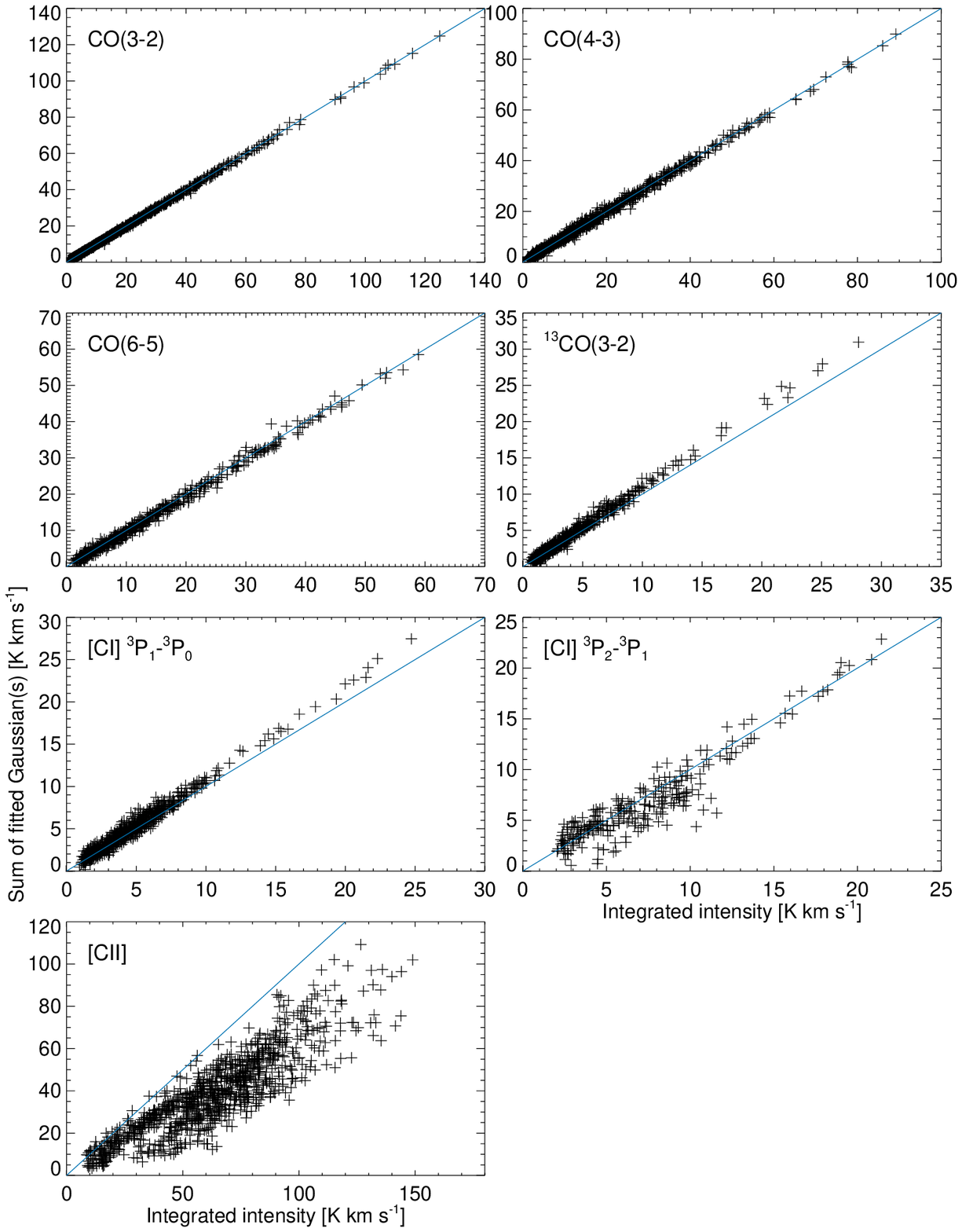}
\caption{The integrated intensity as a sum of the fitted Gaussians vs. the integration over $220$--$250$~\kms.  The blue line corresponds to both values being equal.\label{fig:gaussfit_integ_comp}}
\end{figure}

 Figure~\ref{fig:gaussfit_integ_comp}  shows the relation between the integrated intensity and the sum of the fitted Gaussians for each emission line.  For CO(3-2), both estimates match perfectly, which guarantees that the fitting with two Gaussians is appropriate.  For CO(4-3) and CO(6-5), the integrated intensity is also very well reproduced by the sum of the fitted Gaussians, confirming that the three CO emission lines contain the same two velocity components and give no indication of additional ones.  On the other hand, the plots for \coiso(3-2) and \ci\ \transl\ show that the sum of the fitted Gaussian provides larger values by 5--10\% than the integrated intensity where the emissions are strong.  The \ci\ \transu\ emission may have the same trend for the six strongest  data points, but the S/N of the line is much smaller and the trend is not conclusive.  Looking at individual spectra, we find that this is because the line widths of \coiso(3-2) and \ci\ \transl\  are slightly narrower at these positions than those of CO(3-2).  When we fit the spectra with the width  as a free parameter, but fixing the center and the amplitude to the values that we get in the previous fit, the integrated intensity is well produced in both lines.  This is likely an optical depth effect because the CO(3-2) and CO(4-3) are optically thick at the line center, the optical depth of the CO(6-5) emission at the strongest position is about unity, whereas the \coiso(3-2) and \ci\ emissions are optically thin (see Sect.~\ref{subsec:result_tex}).  In any case, the difference in the line width is small, since the dynamics makes the line width broader than the thermal width.  

The largest difference is for the case of \cii.  The plot for \cii\ in Fig.~\ref{fig:gaussfit_integ_comp} shows a clear trend: the sum of the fitted CO-constrained Gaussians only contributed a fraction of the total flux calculated by the integration over the whole velocity range.  The large scatter in the plot also indicates a large variation in the fraction of the \cii\ flux in the velocity components correlated with the CO and \ci\ emission.  For example, at the center of N159~W (p1, see Figs.~\ref{fig:integ_map} and \ref{fig:spectra}), $\sim 20$\% of the \cii\ integrated line emission cannot be contributed by CO-constrained Gaussians, whereas this ratio is $\sim 40$\% at p7 and more than $50$\% at p9.  The origin for this \cii\ emission may be due to gas ablating from the dense cloud cores, or may be due to additional gas not immediately associated with the dense cores.

We also examined the ratio of the integrated intensity of the different emission lines derived as the sum of the fitted Gaussians to that of CO(3-2).  They show clear spatial variations, which is already visible from the example spectra (Fig.~\ref{fig:spectra}). The \cii/CO(3-2) ratio varies most significantly: more than one order of magnitude across the map.  The ratio is low where the CO(3-2) emission is strong ($0.6$--$1$ for p1, p10, and p14), indicating a low radiation field and/or a high density \citep{Roellig2006,Kaufman2006}, and it increases towards regions where diffuse gas is likely dominant ($\sim 11$ at p7).  A full analysis of these amplitude ratios using PDR modeling will be presented in the follow-up paper (Okada et al. in preparation).

\subsection{Column densities of \cplus, C, and CO} \label{subsec:result_tex}

In this section, we present and discuss the spectral and spatial variations of \cplus, C, and CO column densities and their ratios.  We derive column densities under the simplifying assumption of local thermodynamic equilibrium (LTE) for each species, and a uniform excitation temperature (\tex) for different velocity bins at each position.

\begin{figure*}
\centering
\includegraphics[bb=40 0 424 350,width=0.39\hsize,clip]{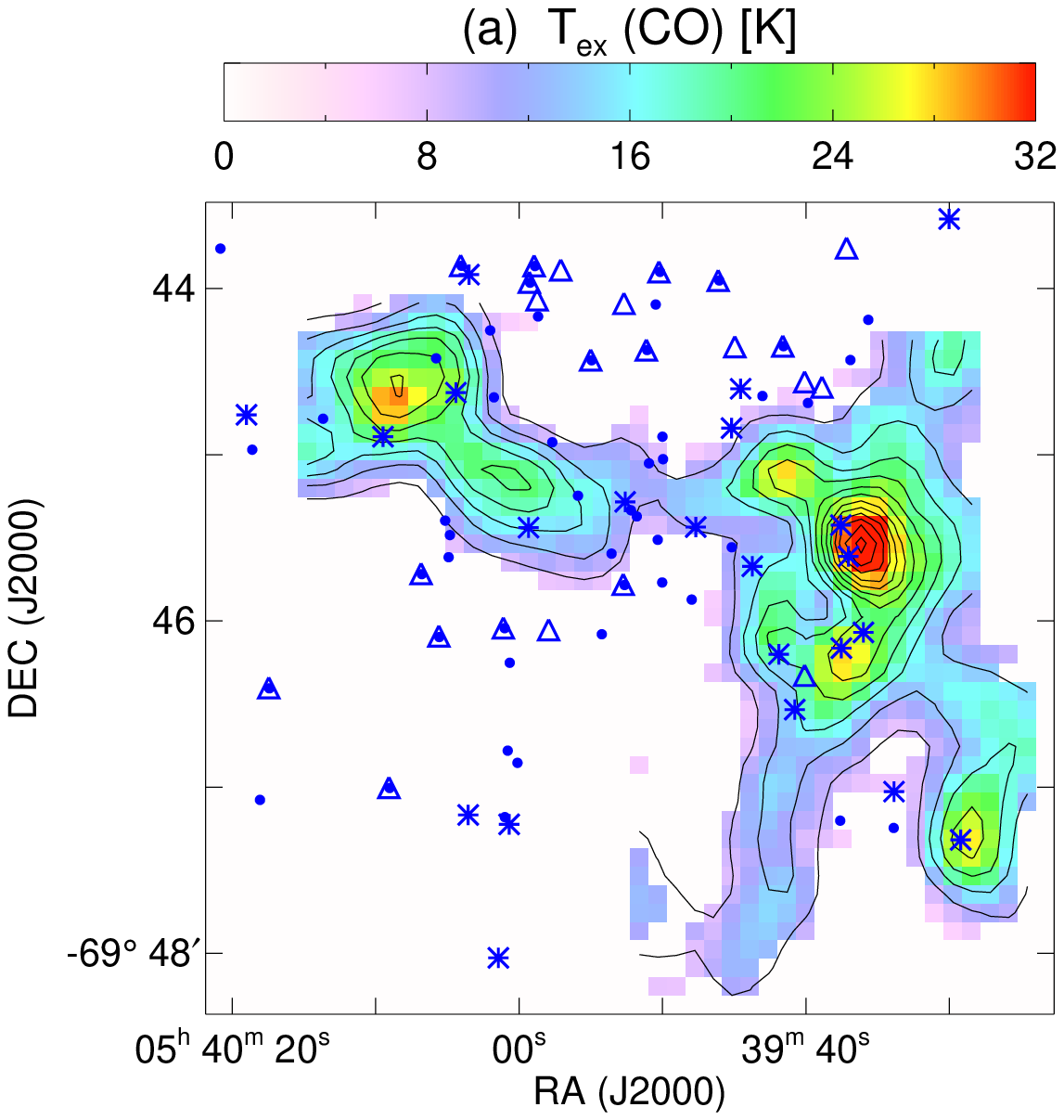}
\hspace*{-0.8cm}
\includegraphics[bb=95 0 424 350,width=0.334\hsize,clip]{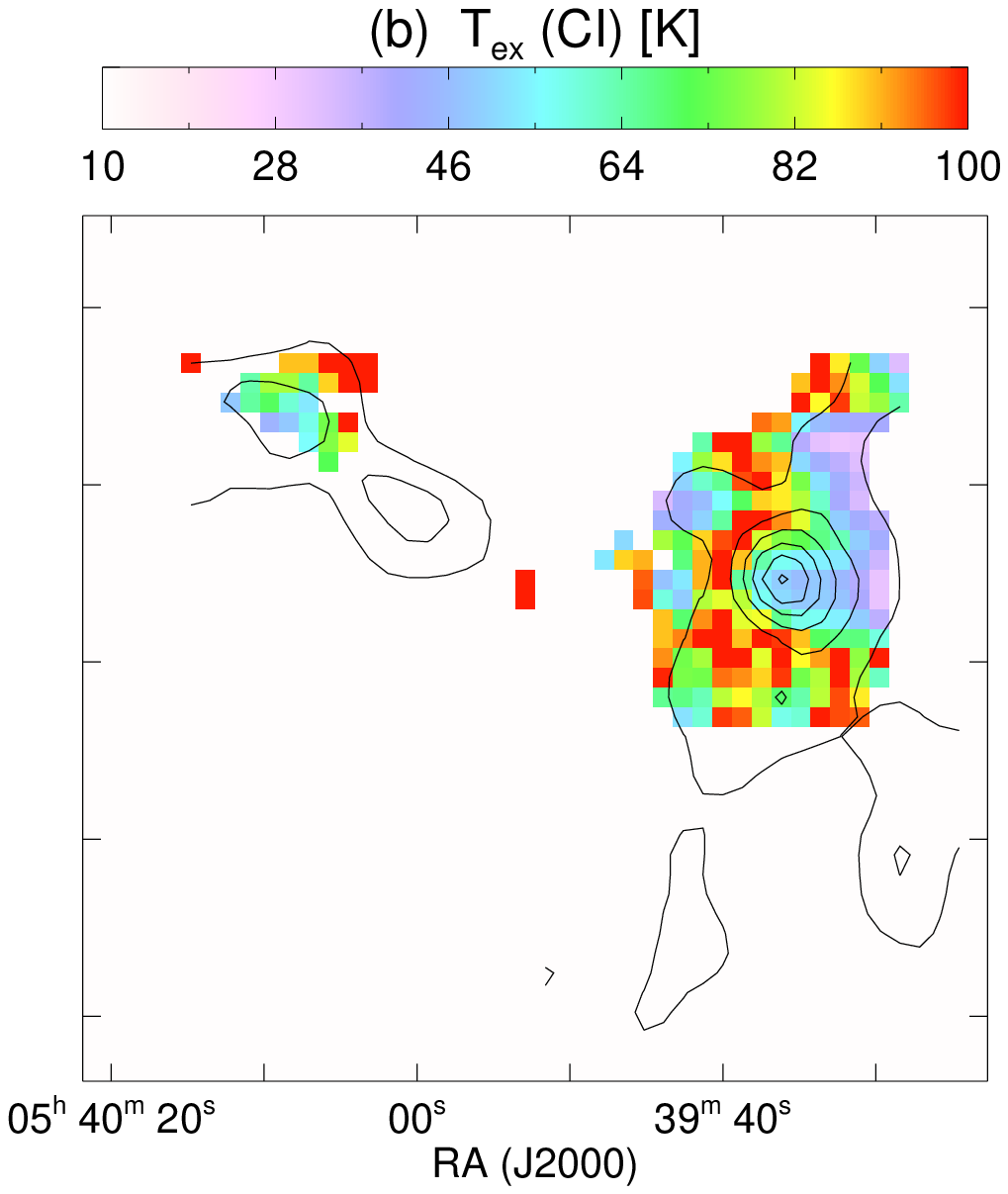}
\hspace*{-0.8cm}
\includegraphics[bb=95 0 424 350,width=0.334\hsize,clip]{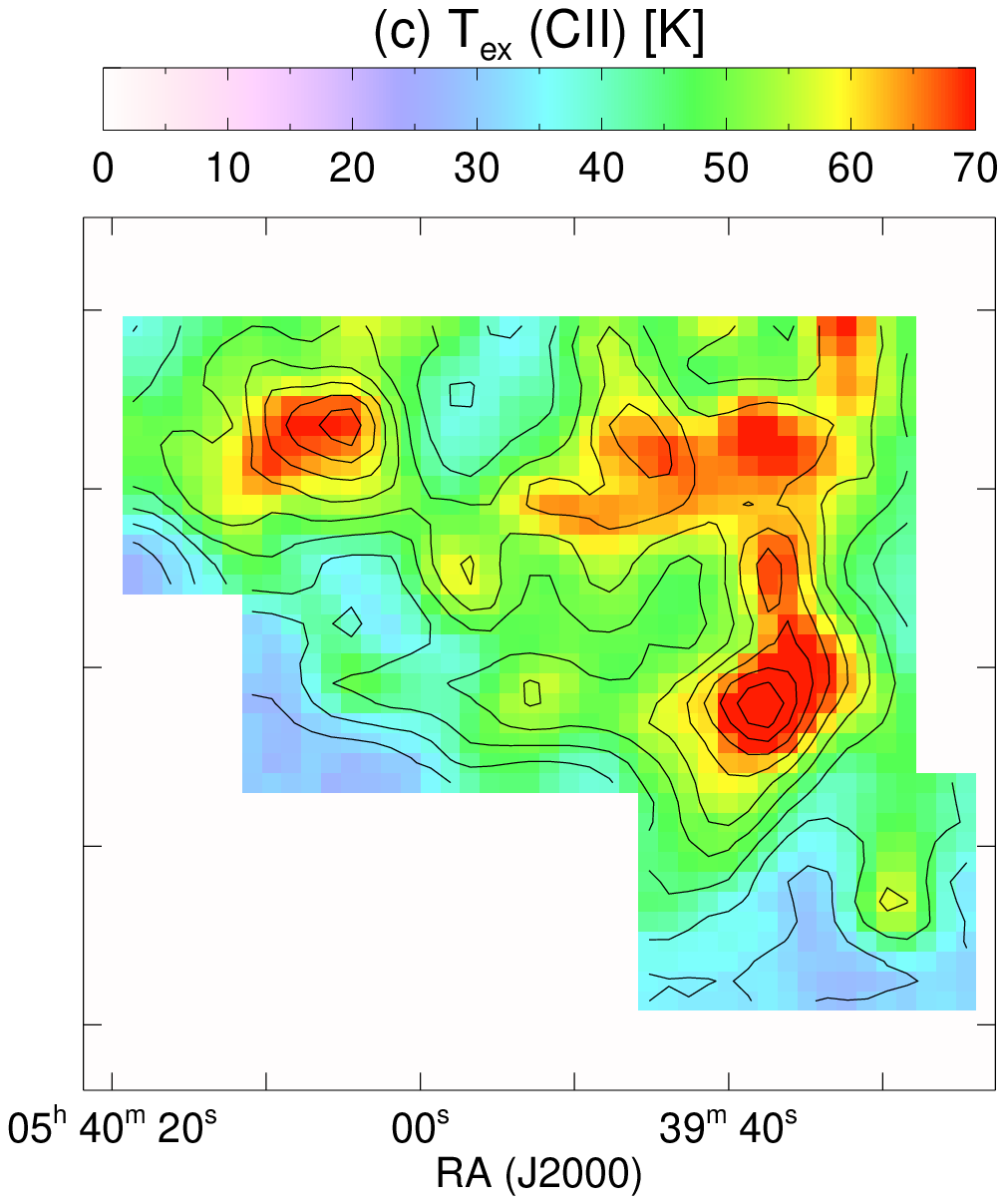}
\caption{(a) CO excitation temperature \tex\ assuming a beam filling factor $\eta=0.5$ overlayed with contours of CO(3-2) integrated intensity.  The positions of YSOs identified by \citet{Chen2010} are marked with asterisks, OB stars (earlier than B4) from \citet{Farina2009} are marked with triangles, and OB star candidates from \citet{Nakajima2005}, taken from \citet{Chen2010}, are marked with circles. (b) \tex\ derived from the two \ci\ lines overlayed with the contours of \ci\ \transl\ integrated intensity. (c) \tex\ derived by assuming $\tau=1$ for the \cii\ emission overlayed with the contours of \cii\ integrated intensity.\label{fig:Tex}}
\end{figure*}

Assuming that CO and \coiso\ have the same \tex\ and beam (area) filling factor ($\eta$), the peak temperature ($T_\mathrm{B}$) ratio of CO(3-2) and \coiso(3-2) is described by $[1-\exp(-\tau_{12})]/[1-\exp(-\tau_{13})]$, where $\tau_{12}$ and $\tau_{13}$ are the peak optical depth of the CO(3-2) and \coiso(3-2) emission, respectively.  We assume an isotope ratio $^{12}$C/$^{13}$C of 49 as derived for the LMC/N113 region by \citet{Wang2009} and use it as $\tau_{12}/\tau_{13}$.  We derive $\tau_{13}$ by solving Eq.~(\ref{eq:tau13}) numerically pixel by pixel,
\begin{equation}
\frac{T_\mathrm{B}\textrm{(CO(3-2))}}{T_\mathrm{B}(^{13}\textrm{CO(3-2))}} = \frac{1-\exp(-49\times\tau_{13})}{1-\exp(-\tau_{13})}\label{eq:tau13}
;\end{equation}
$\tau_{13}$ varies across the map from 0.06 to 0.5, which corresponds to $\tau_{12}=3$--35, i.e., CO(3-2) emission is optically thick everywhere.  In general, \tex\ can be expressed as
\begin{equation}
T_\mathrm{ex}=\frac{h\nu}{k}\left[\ln\left\{\left(1-e^{-\tau}\right)\eta\frac{h\nu}{kT_\mathrm{B}}+1\right\}\right]^{-1}. \label{eq:tex}
\end{equation}
Here we use $\tau_{13}$ and $T_\mathrm{B}$(\coiso(3-2)) to derive \tex\ of CO.  The resulting \tex\ for the case of $\eta=0.5$ is shown in Fig.~\ref{fig:Tex}a.  In the figure, the positions of early-type stars (earlier than B4) from \citet{Farina2009}, YSOs from \citet{Chen2010}, and OB star candidates from \citet{Nakajima2005} \citep[taken from][]{Chen2010} are overplotted.  Most of the cores are associated with one or more of these, which is consistent with a higher \tex\ at those positions.  The calculation with $\eta=1$ gives a lower limit of \tex, and the \tex\ at N159~W is 21~K, 36~K, and 150~K for $\eta=1$, $0.5$, and $0.1$, respectively.  \citet{Galametz2013} derive a dust temperature from the far-infrared continuum emission with 36\arcsec\ resolution, which is 30--35~K in our observed region.  When we estimate \tex\ at N159~W and E in the same spatial resolution, $\eta=0.3$--0.4 gives the same temperature for \tex\ as their dust temperature.  \citet{Minamidani2011} derive the kinetic temperature from a LVG analysis with 45\arcsec\ resolution to be 40--82~K and 65--151~K in N159~W and E, respectively.  If we assume that \tex\ equals their kinetic temperature, it requires a value for $\eta$ of $<0.3$.  The derived range of filling factors is consistent with the reasonable scenario, that clumps with an enhanced density by more than one or two orders of magnitude fill $\sim 60$\% or $\sim 35$\% of the projected area, respectively.  In the following, we use the results for $\eta=0.5$, together with $\eta=0.1$ and $0.2$ for some cases.

The optical depth of an emission line $\tau_\nu$ is related to the column density of the lower state $N_l$ by 
\begin{equation}
\tau_\nu=\frac{c^2}{8\pi}\frac{1}{\nu^2_0}\frac{g_u}{g_l}N_lA_{ul}\left[1-\exp\left(-\frac{h\nu_0}{kT_\mathrm{ex}}\right)\right]\phi_\nu\label{eq:tau}
,\end{equation}
where subscripts $u$ and $l$ indicate the upper and lower state, $g$ is the statistical weight, $A_{ul}$ is the Einstein coefficient, and $\phi_\nu$ is the normalized line profile \citep{ToolsOfRadioAstronomy}.  Under the LTE assumption
\begin{equation}
\frac{N_l}{N_\mathrm{tot}}=\frac{g_l}{Z}\exp\left(-\frac{E_l}{kT_\mathrm{ex}}\right)
,\end{equation}
where $N_\mathrm{tot}$ is the total column density, $E_l$ is the energy of the lower state, and $Z$ is the partition function
\begin{equation}
Z=\sum_{l=0}^\infty g_l \exp\left(-\frac{E_l}{kT}\right)
.\end{equation}
Therefore, the column density in each velocity bin ($dN_v$) can be derived as
\begin{equation}
dN_v=\tau_V\times\frac{8\pi\nu^2_0}{c^2}\frac{Z}{g_u}\frac{1}{A_{ul}}\exp\left(\frac{E_l}{kT_\textrm{ex}}\right)\left[1-\exp\left(-\frac{h\nu_0}{kT_\textrm{ex}}\right)\right]^{-1} dv\label{eq:dN_v}
;\end{equation}
$\tau_v$ is derived from the brightness temperature at each velocity bin ($T_\mathrm{B}(v)$) as
\begin{equation}
\tau_v=-\ln\left[1-\frac{kT_\mathrm{B}(v)}{\eta h\nu}\left\{\exp\left(\frac{h\nu}{kT_\mathrm{ex}}\right)-1\right\}\right] \label{eq:tau_v}
.\end{equation}

\begin{figure}
\centering
% printer
\includegraphics[bb=40 0 424 350,width=0.572\hsize,clip]{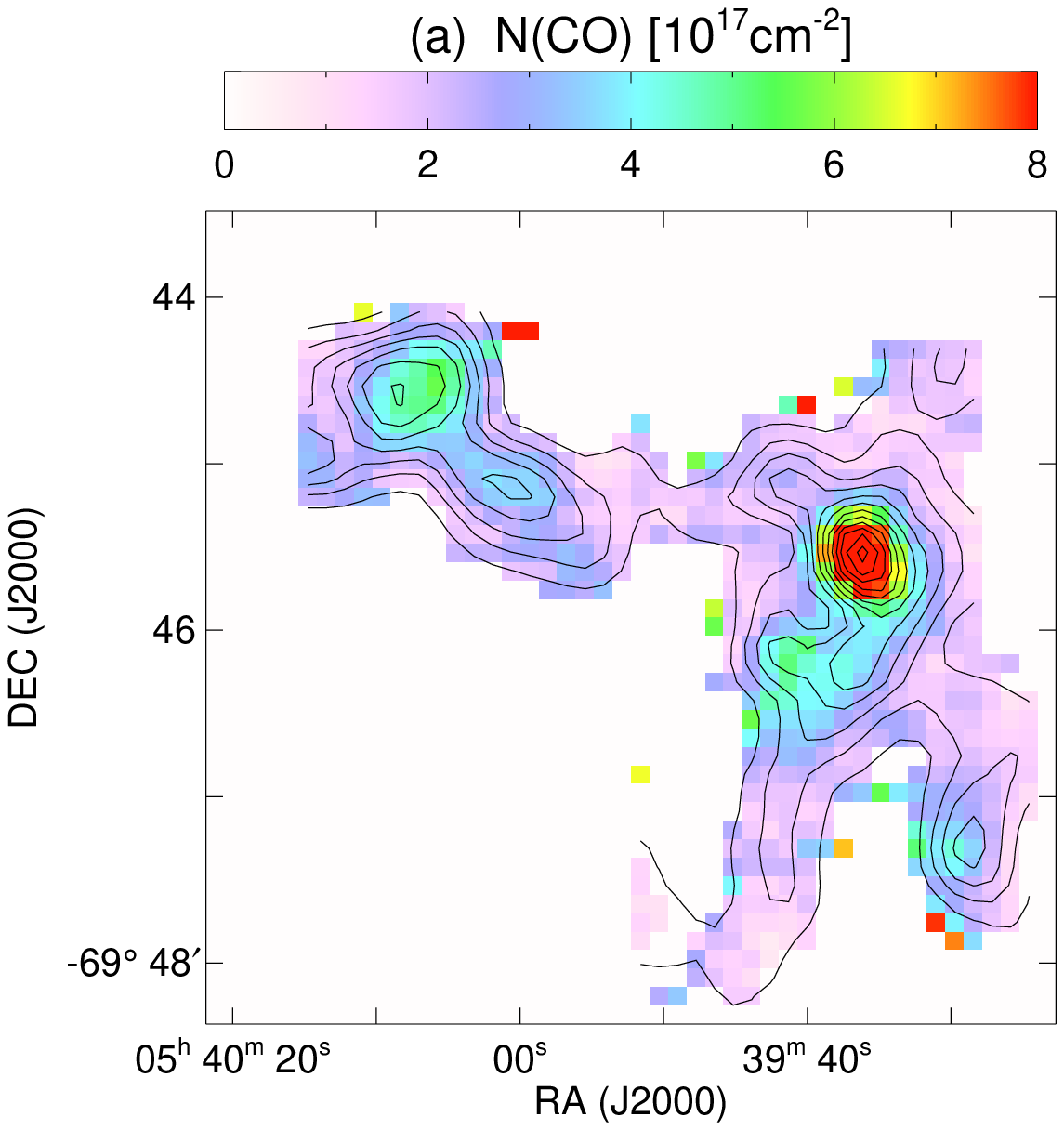}
\hspace*{-0.8cm}
\includegraphics[bb=95 0 424 350,width=0.49\hsize,clip]{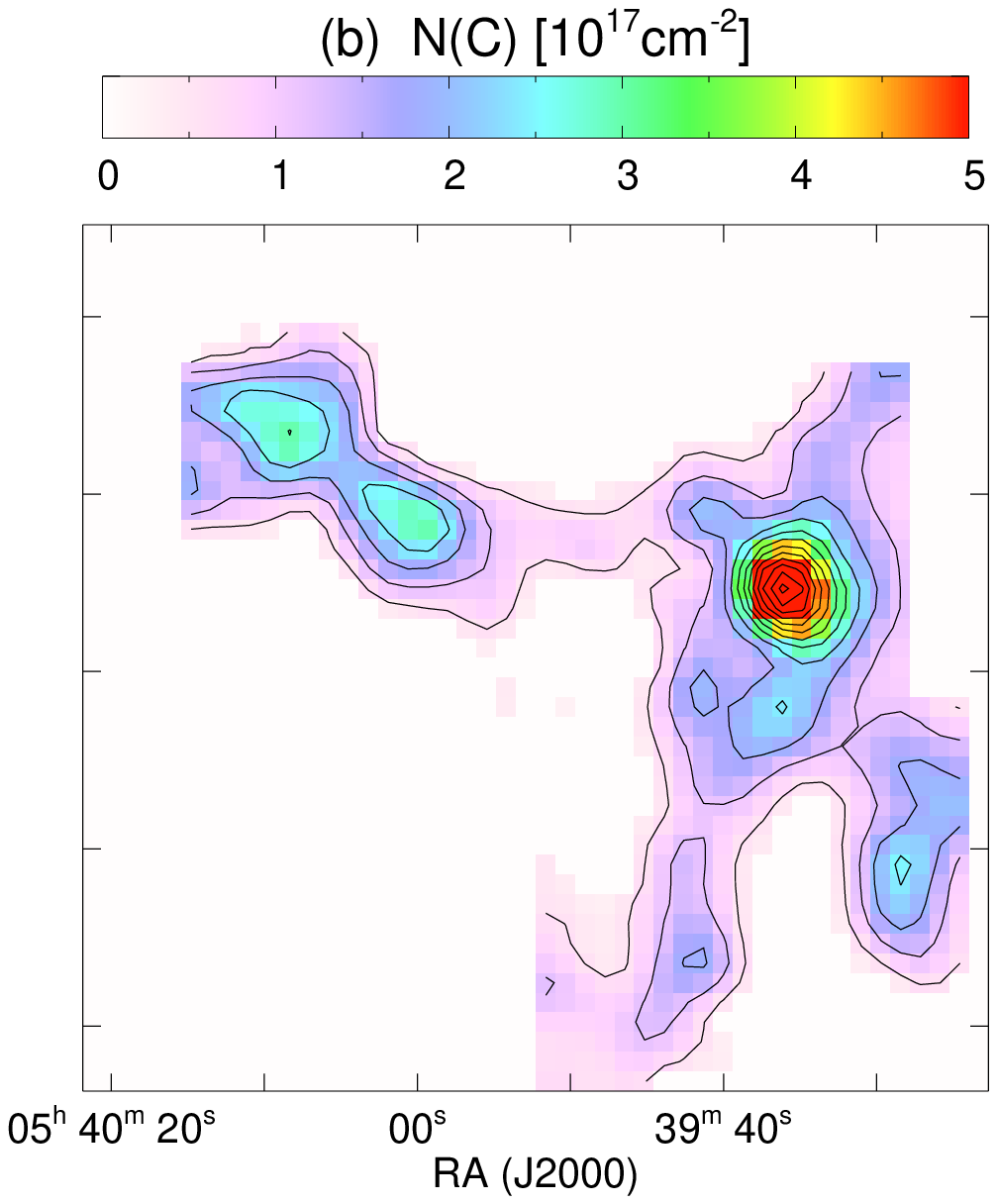}
\includegraphics[bb=40 0 424 350,width=0.572\hsize,clip]{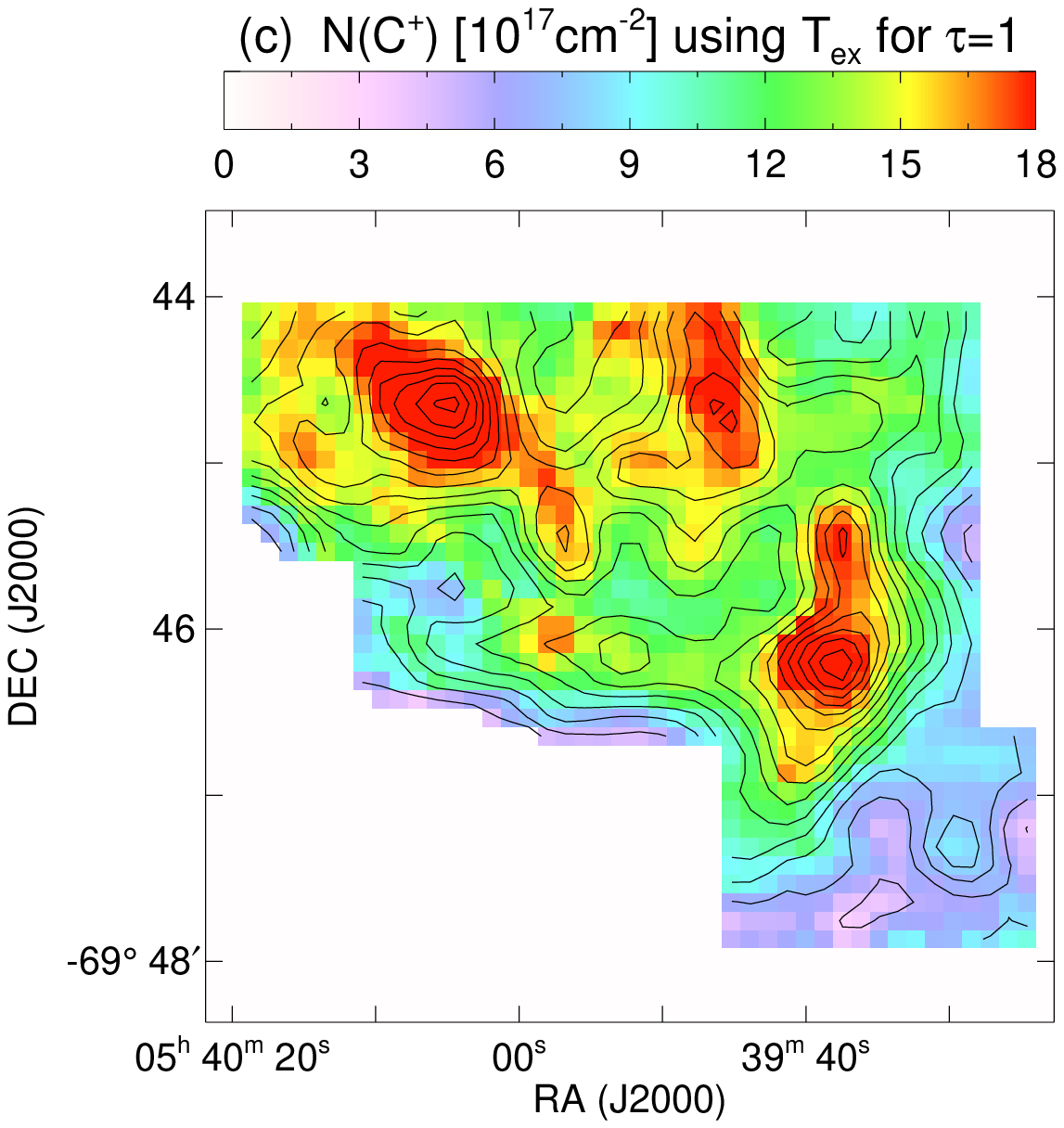}
\hspace*{-0.8cm}
\includegraphics[bb=95 0 424 350,width=0.49\hsize,clip]{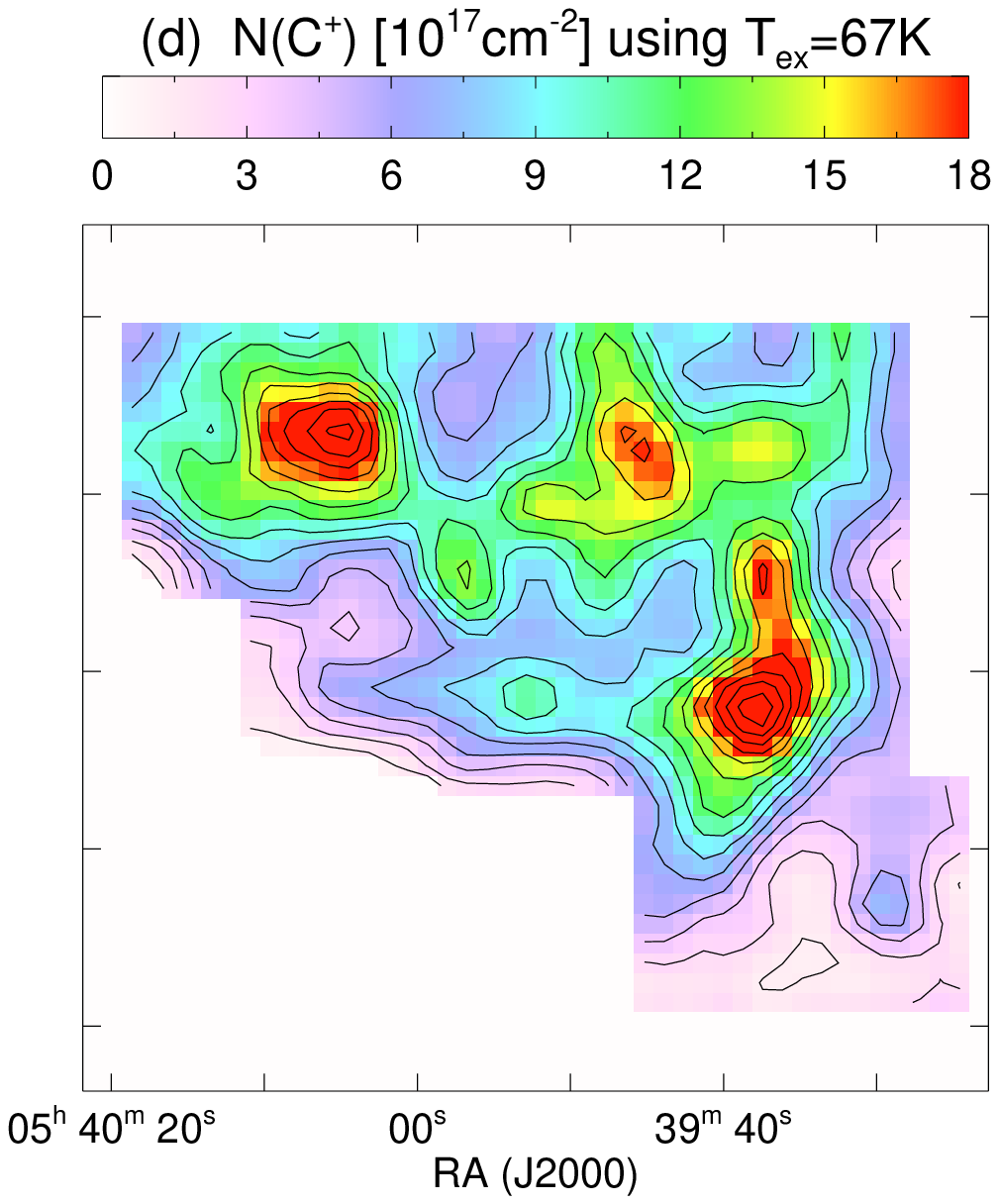}
\caption{The total column density of CO, C, and \cplus\ for the $\eta=0.5$ case overlayed with the contours of the integrated intensity of CO(3-2) in (a), \ci\ \transl\ in (b), and \cii\ in (c) and (d).  $N$(C) is derived by assuming \tex$=74$~K, and for $N$(\cplus), both cases of $\tau=1$ (c) and \tex$=67$~K (d) are shown (see text).\label{fig:column}}
\end{figure}

In the case of the CO $J+1\rightarrow J$ transition, $E_l=hBJ(J+1)$, $g_u=2(J+1)+1$, $Z=kT_\mathrm{ex}/hB$, and $A_{ul}=64\pi^4/(3hc^3)\times \nu_0^3\mu^2\times (J+1)/(2J+3)$ \citep{ToolsOfRadioAstronomy}, therefore 
\begin{equation}
\begin{split}
dN_v(\mathrm{CO})=\tau_V\times\frac{3ckT_\mathrm{ex}}{8\pi^3\nu_0 B\mu^2}&\frac{1}{J+1}\exp\left(\frac{hBJ(J+1)}{kT_\mathrm{ex}}\right)\\
&\times\left[1-\exp\left(-\frac{h\nu_0}{kT_\mathrm{ex}}\right)\right]^{-1}dv
\end{split}
,\end{equation}
where $\mu$ is the dipole moment and $B$ is the rotational constant.  We used $T_\mathrm{B}(v)$ of \coiso(3-2) and derived $dN_v$(\coiso) and converted it to $dN_v$(CO) by multiplying $^{12}$C/$^{13}$C$=49$.  The total column density $N$(CO) was derived by integrating $dN_v$(CO) over the frequency range corresponding on the velocity of 220--250~\kms\ (Fig.~\ref{fig:column}).  The spatial distribution of $N$(CO) follows the integrated intensity of CO(3-2) and \tex\ (Fig.~\ref{fig:Tex}a).

\begin{figure}
\centering
\includegraphics[width=\hsize]{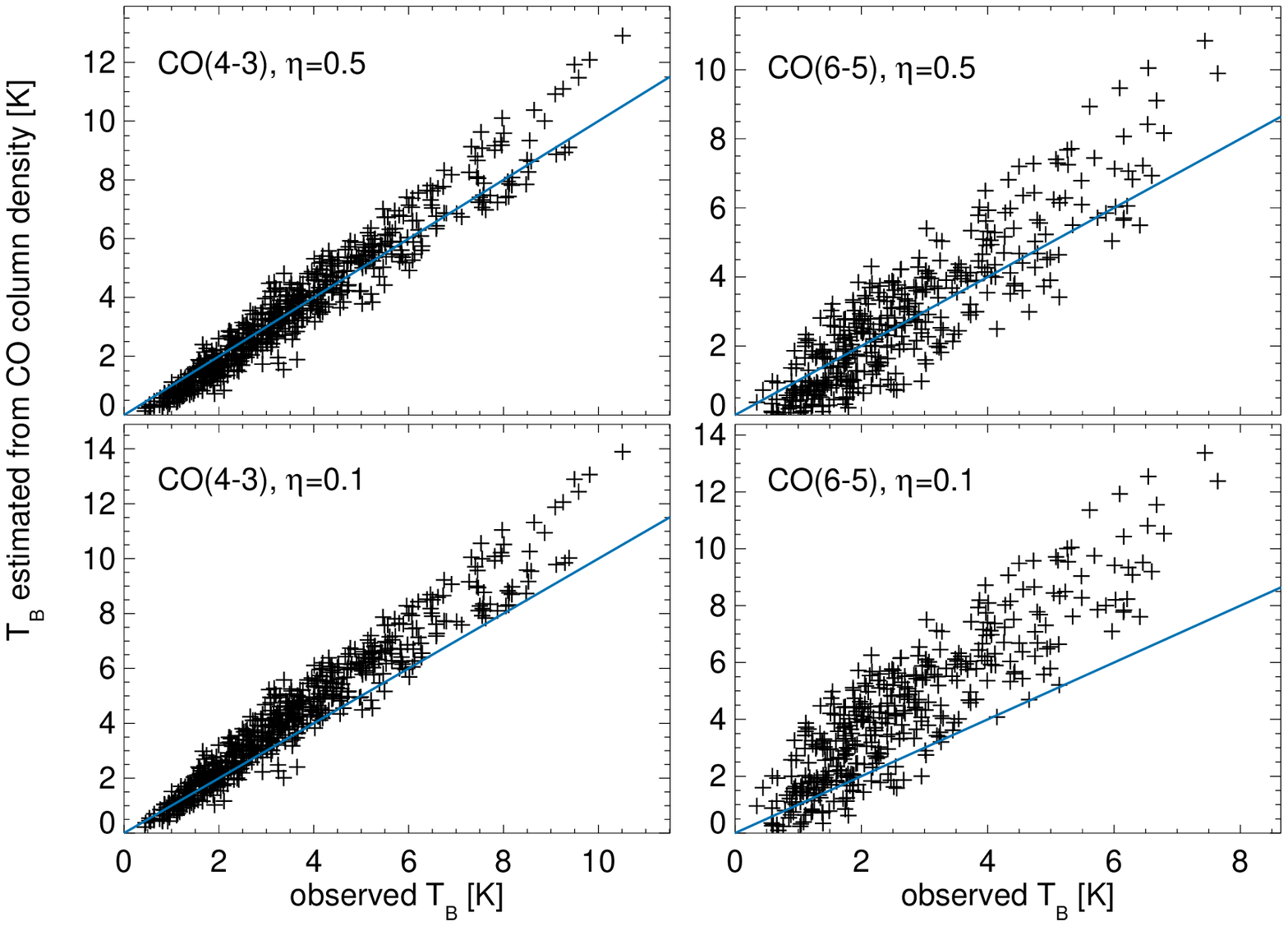}
\caption{Peak brightness temperature of CO(4-3) (left panels) and CO(6-5) (right panels) estimated from the CO column density derived by using \tex\ under the assumption of $\eta=0.5$ (upper panels) and 0.1 (lower panels) for all CO lines against observed $T_B$ (20\arcsec\ resolution).  Blue lines indicate the case where the estimated and observed $T_B$ is the same.\label{fig:comp_TB_co}}
\end{figure}

In Fig.~\ref{fig:comp_TB_co} we compare the observed peak temperatures $T_B$ of CO(4-3) and CO(6-5) with the expected ones from the peak (in terms of velocity) of $dN_v$(CO) assuming that all transitions have the same \tex\ and $\eta$.  In the case of $\eta=0.5$, the observed $T_B$ is well reproduced, indicating that the assumption of identical \tex\ and $\eta$ for all transitions is justified.  When $\eta=0.1$, the observed $T_B$ of CO(4-3) and CO(6-5) is systematically higher by $\sim 20$\% and $\sim 60$\%, respectively.  The fact that the observed $T_B$ of CO(6-5) is weaker is likely due to CO(6-5) not being thermalized.  The critical densities of CO(3-2) and CO(6-5) are (3--4)$\times 10^4$~\cc\ and $3\times 10^5$~\cc, respectively, for 10--100~K \citep[calculated using][]{Flower2001,Goorvitch1994}.  \citet{Minamidani2011} derived $n($\hh) of (4--5)$\times 10^3$~\cc\ for N159~W and E, and \citet{Pineda2008} discuss the result of PDR modeling with an average density of the clump ensemble of about $10^5$~\cc.  These densities are consistent with the assumption that the CO(6-5) emission does not reach LTE, in contrast to CO(3-2).  Calculations with the RADEX radiative transfer program show that the CO(3-2)/CO(6-5) intensity ratio at 30--100~K changes by more than a factor of  2 between $10^4$--$10^5$~\cc\ \citep{vanderTak2007}, which can explain the CO(6-5) discrepancy in Fig.~\ref{fig:comp_TB_co} in this work.  Alternatively, it is also possible that $\eta$ of CO(6-5) is smaller than that of CO(3-2) because it traces warmer gas which is expected to be in a thinner  layer.

At positions where both \ci\ lines are detected, we derive the \ci\ excitation temperature \tex\ and the peak optical depth $\tau$ from their ratio and their peak temperature.  The optical depth ratio of both transitions in \ci\ can be expressed in terms of \tex\ under the LTE assumption (Eq.~\ref{eq:tau}).  Combining with Eq.~\ref{eq:tex} for both transitions provides two independent equations in terms of \tex, $\eta$, and $\tau_{10}$ (the optical depth of \transl) as unknown variables, and $T_B$ as observed quantities.  Since the S/N of \ci\transu\ is low, and the line profiles of the two \ci\ lines are the same with the exception of a few positions mentioned in Sect.~\ref{subsec:result_structure}, we did not use directly $T_B$(\transl) and $T_B$(\transu) as observed variables, but used $T_B$(\transl) and $I$(\transu)/$I$(\transl), where $I$ is the integrated intensity.  We calculated $\tau_{10}$ and \tex\ numerically for individual $\eta$.  The resultant $\tau_{10}$ is less than 1 at all pixels for all $\eta$, and \tex\ is insensitive to $\eta$.  For the fully optically thin case, $I$(\transu)/$I$(\transl) is proportional to $\tau_{21}/\tau_{10}$, therefore independent on $\eta$.  The median of \tex\ across the map is 70--80~K for $\eta\geq 0.2$ and 90~K for $\eta=0.1$.  In Fig.~\ref{fig:Tex}b the spatial distribution of \tex\ for $\eta=0.5$ is shown.  In contrast to \tex\ derived from the CO emissions, \tex\ for \ci\ is lower at the N159~W core than the surroundings.  If we assume that \tex\ of CO and \ci\ are similar (within 10~K), the comparison at p1 (N159~W), p4, and p10 (N159~E) gives $\eta$ of 0.3--0.4, $<0.2$, and 0.1--0.2, respectively.

To derive the total neutral atomic carbon column density from Eqs.~(\ref{eq:dN_v}) and (\ref{eq:tau_v}), we sum over the three fine structure levels
\begin{equation}
Z=g_0+g_1\exp\left(-\frac{E_1}{kT_\mathrm{ex}}\right)+g_2\exp\left(-\frac{E_2}{kT_\mathrm{ex}}\right) \label{eq:Z}
\end{equation}

Since the region where the \ci\ \transu\ emission is detected is small and the uncertainty of the derived temperature is relatively large because of the weak \ci\ \transu\ emission, we examine both cases using the \tex\ derived from the \ci\ ratio and a constant value of its median to estimate the column density.  The derived column densities match within 10\%, even less than 5\% at most positions.  In the following, we use the case with constant \tex\ (Fig.~\ref{fig:column}) because it covers a larger spatial area.

Since there is only a single \cii\ line, we need to assume its optical depth in order to estimate \tex\ using Eq. (\ref{eq:tex}), or assume a constant \tex\ for the column density derivation.  In several Galactic PDRs, the optical depth of the \cii\ line is estimated to be (0.6--3.2) using the hyperfine components of the \ciiiso\ emission \citep{Ossenkopf2013,Graf2012,Stacey1991}.  In the N159 region, the strongest \ciiiso\ hyperfine component is blended with the broad main isotope line, and the upper limit of the second strongest \ciiiso\ hyperfine component does not give a useful constraint to the optical depth of \cii.  Figure~\ref{fig:Tex}c shows the estimate of \tex\ for a constant optical depth $\tau=1$ \citep[see][]{Kaufman1999} for the peak velocity and $\eta=0.5$.  The spatial distribution follows the distribution of the \cii\ brightness because a stronger peak temperature is attributed to a higher temperature.  When we assume $\tau=0.5$, \tex\ increases by 15--37\%.  For the optically thin case, $\tau$ is coupled with $\eta$.  When we assume $\tau=3$ and $\tau\gg 1$, \tex\ goes down typically by 10--16\% and 11--18\%, respectively.  The latter case gives the lower limit of \tex.  

The column density of \cplus\ is derived using Eqs.~(\ref{eq:dN_v}) and (\ref{eq:tau_v}), together with the first two terms in Eq.~(\ref{eq:Z}) because \cplus\ is a two-level system.  We examine both cases using the \tex\ derived by assuming $\tau=1$ and a constant value (Figs.~\ref{fig:column}c and d).  The constant value is determined, depending on $\eta$, as the largest value of \tex\ estimated from the $\tau\gg 1$ case across the map, which is 67~K and 199~K for $\eta=0.5$ and 0.1, respectively.  In this way, the constant \tex\ is above the \tex\ lower limit at any position of the map.  We then estimate $\tau$ in case of this constant \tex.  It is fully optically thick at the \cplus\ peak of p4, and regions around other peaks (N159~W, E, p7) have $\tau\sim 1$ at the peak velocity.  For the constant \tex\ case, the contrast of the \cii\ emission is attributed to the difference in $\tau$; therefore, the derived column density basically reproduces the \cii\ intensity distribution (Fig.~\ref{fig:column}d).  For the constant $\tau$ case, the stronger \cii\ emission is attributed to a higher \tex\ (Fig.~\ref{fig:Tex}c), but the column density is insensitive to \tex.  Thus, the large variation of the \cii\ velocity profile causes a different $\tau_\nu$ profile from position to position and results in a different spatial distribution of $N$(\cplus) from the \tex\ and the integrated intensity distribution.  The higher $N$(\cplus) in the eastern part of the map (Fig.~\ref{fig:column}c) is due to a wider line profile there because the constant $\tau$ is at the peak of the line and a wider line makes the integrated column density larger.  Qualitatively, positions with strong \cii\ emission are the outer part of PDRs where the temperature is higher and the column density of \cplus\ is higher than toward the molecular core.  Therefore, the assumptions of constant $\tau$\ and constant \tex\ are the two extreme cases, and the reality should be somewhere in between.  In the following analysis we use the constant \tex\ case because it is unlikely that the optical depth is the same between cores or the \cii\ peak positions and regions in between where the \cii\ emission is much fainter.

\begin{table}
\caption{Total column density of \cplus, C, and CO, and $N_\mathrm{H}$ estimated from $N$(\cplus)$+N$(C)$+N$(CO).\label{table:NH}}
\centering
\begin{tabular}{ccccc}
\hline\hline
Pos. & $N$(\cplus) & $N$(C) & $N$(CO) & $N_\textrm{H}$\\
 & [$10^{17}$\,cm$^{-2}$] & [$10^{17}$\,cm$^{-2}$] & [$10^{17}$\,cm$^{-2}$] & [$10^{22}$\,cm$^{-2}$] \\
\hline
\multicolumn{5}{c}{$\eta=0.1$}\\
\hline
 1 &   58 &   44 &  139 &  30 \\
 4 &  131 &   12 &   35 &  22 \\
 8 &   52 &    3 &    4 &   7 \\
10 &   65 &   14 &   35 &  14 \\
11 &   83 &    9 &   28 &  15 \\
\hline
\multicolumn{5}{c}{$\eta=0.2$}\\
\hline
 1 &   32 &   19 &   44 &  12 \\
 4 &   66 &    5 &   12 &  10 \\
 8 &   28 &    1 &    2 &   4 \\
10 &   36 &    6 &   12 &   7 \\
11 &   45 &    4 &   10 &   7 \\
\hline
\multicolumn{5}{c}{$\eta=0.5$}\\
\hline
 1 &   16 &    7 &   13 &   4 \\
 4 &   35 &    2 &    4 &   5 \\
 8 &   14 & 0.6 &    1 &   2 \\
10 &   18 &    2 &    4 &   3 \\
11 &   23 &    1 &    4 &   3 \\
\hline
\end{tabular}
\end{table}

\begin{figure*}
\centering
\includegraphics[width=\hsize]{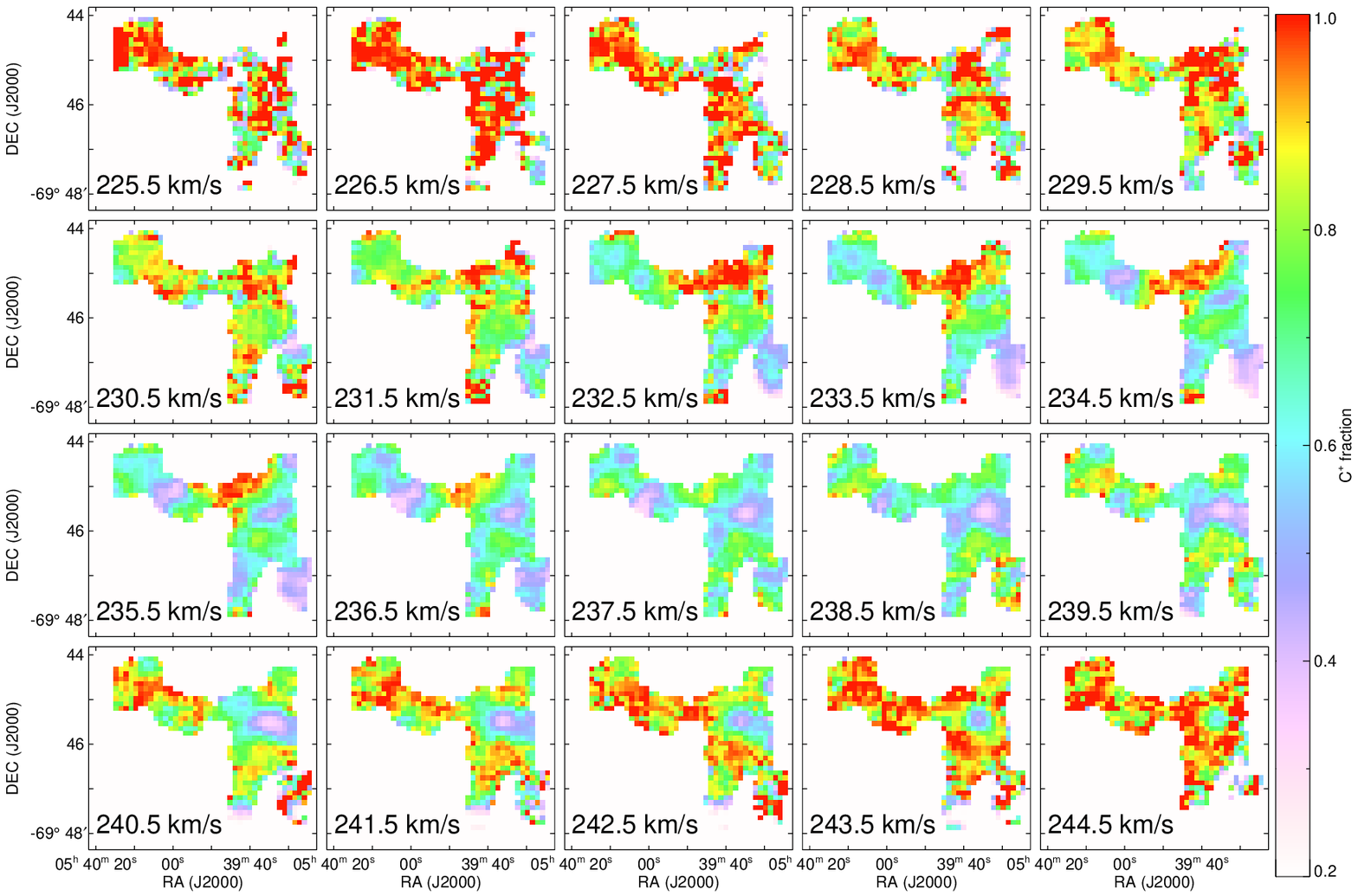}
\caption{Channel map of the fraction of \cplus\ column density against the total carbon column density ($dN_v$(\cplus)$/(dN_v($\cplus$)+dN_v$(C)$+dN_v$(CO)$)$) using the \cplus\ column density derived from \tex$=67$~K and assuming $\eta=0.5$.  The velocity bin is 1~\kms, and the central velocity is shown in each panel.\label{fig:cii_column_channel_const}}
\end{figure*}

\begin{figure*}
\centering
\includegraphics[width=\hsize]{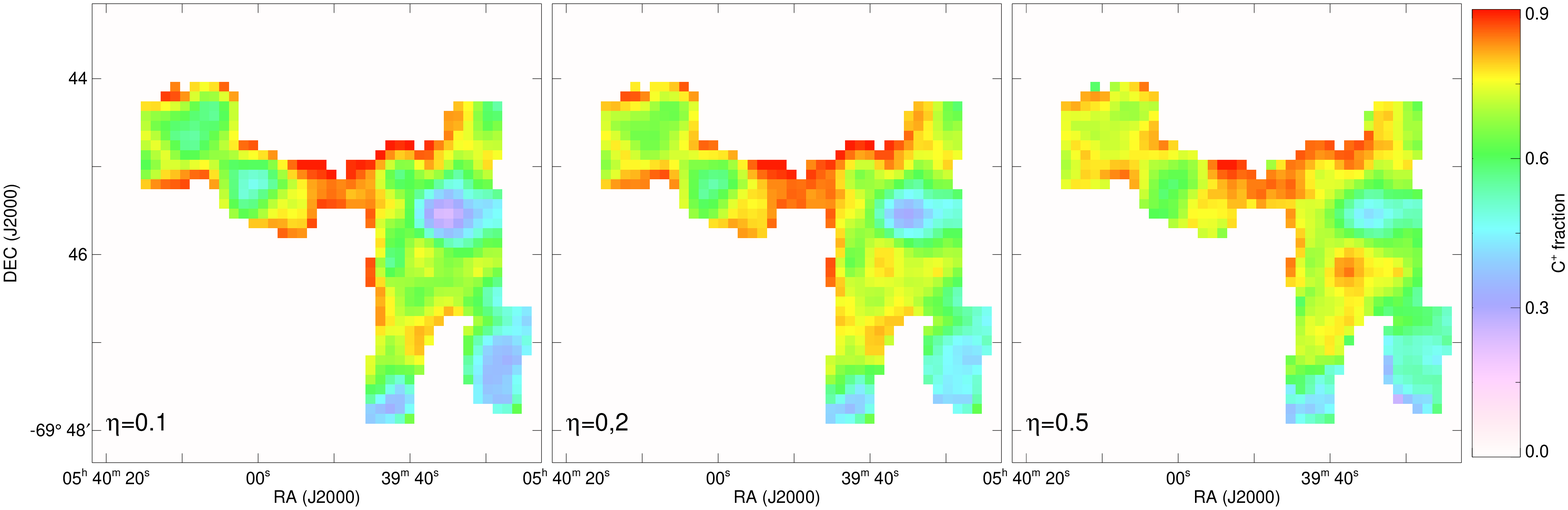}
\caption{The fraction of total column density of \cplus, i.e., $N$(\cplus)$/$($N$(\cplus)$+N$(C)$+N$(CO)), for different $\eta$.\label{fig:cii_totalcolumnfrac}}
\end{figure*}

\begin{figure}
\centering
\includegraphics[width=0.6\hsize]{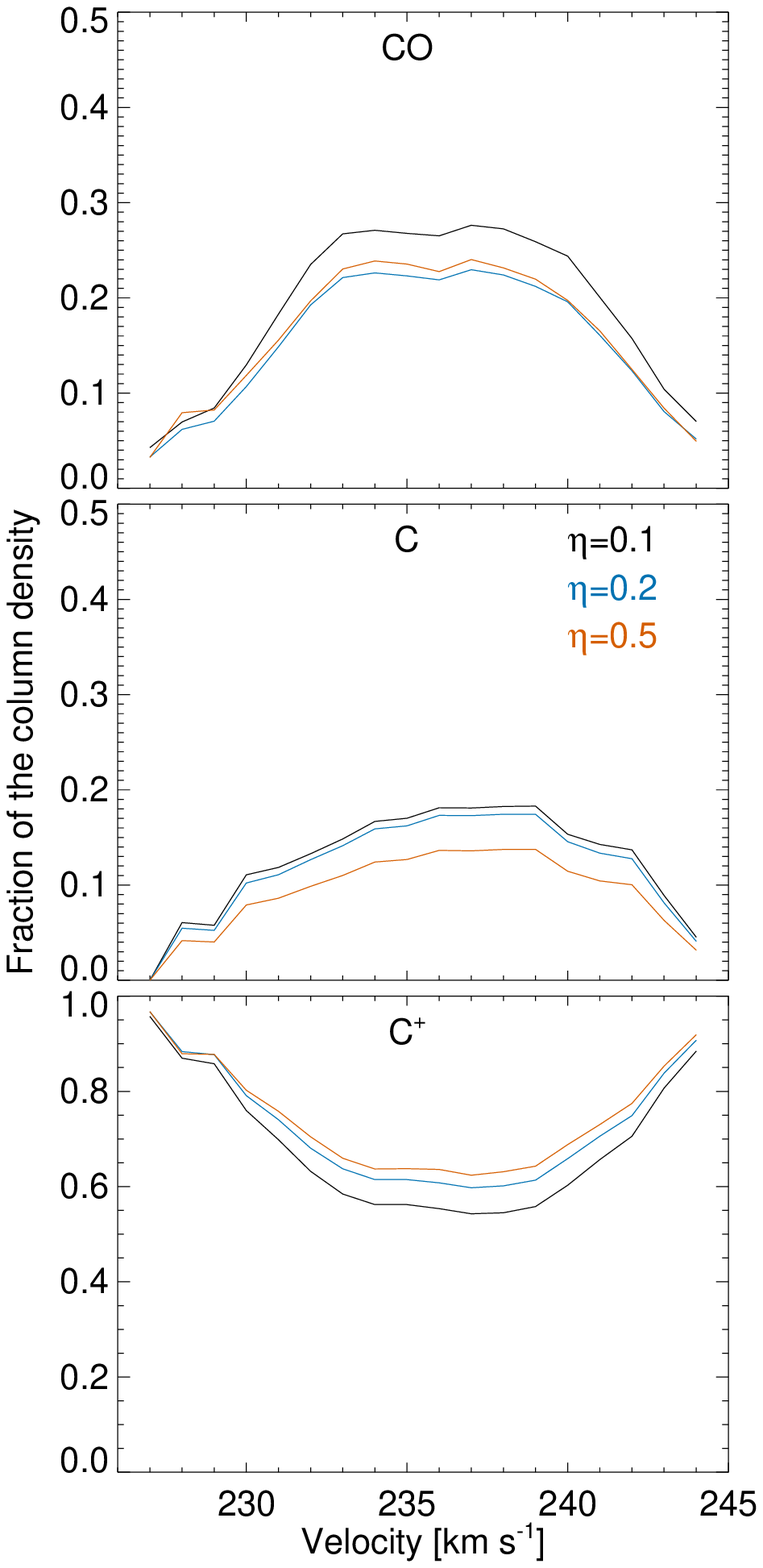}
\caption{Spectra of the fraction of CO, C, and \cplus\ column density against the total column density, i.e., $dN_v$(\cplus) or $dN_v$(C) or $dN_v$(CO) over ($dN_v$(\cplus)$+dN_v$(C)$+dN_v$(CO)) averaged over the map.\label{fig:columnfrac_spec}}
\end{figure}

Assuming that $N$(\cplus)$+N$(C)$+N$(CO) is the total column density of the carbon species in the gas phase, we estimate $N_\mathrm{H}$ using a carbon abundance of $7.9\times 10^{-5}$ \citep{Garnett1999} for selected positions and $\eta$ (Table~\ref{table:NH}).  For a hydrogen density of $10^4$--$10^5$~\cc\ (see above), the depth of the cloud is $0.3$--$3$~pc for $10^{23}$~cm$^{-2}$.  The typical projected size of clumps in the observed region is 30\arcsec, which corresponds to 7~pc at a distance of 50~kpc.  Assuming that the projected size is on the same order as the depth of clouds, a smaller $\eta$ would be preferred.  The total mass, obtained by integrating over the whole observed area, is $2.2\times 10^5 M_\odot$, $4.4\times 10^5 M_\odot$, and $8.5\times 10^5 M_\odot$ for $\eta=0.5$, $0.2$, and $0.1$, respectively.

With $dN_v$ of each species, we estimate the fraction of each species in each velocity bin.  Figure~\ref{fig:cii_column_channel_const} shows the channel map of the \cplus\ column density fraction against the total column density of the three species, i.e., $dN_v$(\cplus)$/$($dN_v$(\cplus)$+dN_v$(C)$+dN_v$(CO)), in the case of $\eta=0.5$ and the \cplus\ column density derived from the constant \tex\ of 67~K.  There are two prominent features here.  One is given by dips in the \cplus\ column density fraction at the N159~W core ($\sim 240$~\kms), N159~E ($\sim 233$~\kms), and southwest of N159~E ($\sim 235$~\kms).  The \cplus\ fraction drops to 30\%, 50\%, and 40\%, respectively, and the rest is shared by \ci\ (30\%, 20\%, 20\%) and CO (40\%, 30\%, 40\%).  On the other hand, \cplus\ is dominant at all velocity bins in the region between N159~W and E.  The second feature is that \cplus\ is clearly dominant at almost all positions in the map for velocities far from the line center ($\lesssim 230$~\kms\ and $\gtrsim 240$~\kms).  The total column density depends on the assumed $\eta$, but the fraction of the column density of each species is less sensitive to $\eta$, and this spatial and spectral trend is largely independent of the value assumed for $\eta$.  Figure~\ref{fig:cii_totalcolumnfrac} shows the fraction of the total \cplus\ against total carbon column density, namely $N$(\cplus)$/$($N$(\cplus)$+N$(C)$+N$(CO)) for different values of $\eta$, confirming the spatial trend, and Fig.~\ref{fig:columnfrac_spec} shows the spectra of the column density fraction averaged over the map, which clearly demonstrates that \cplus\ is the dominant species at velocities of below 230~\kms\ and above 240~\kms.

\subsection{Estimate of the \cii\ emission coming from ionized gas}

\begin{figure}
\centering
\includegraphics[bb=40 0 424 350,width=\hsize,clip]{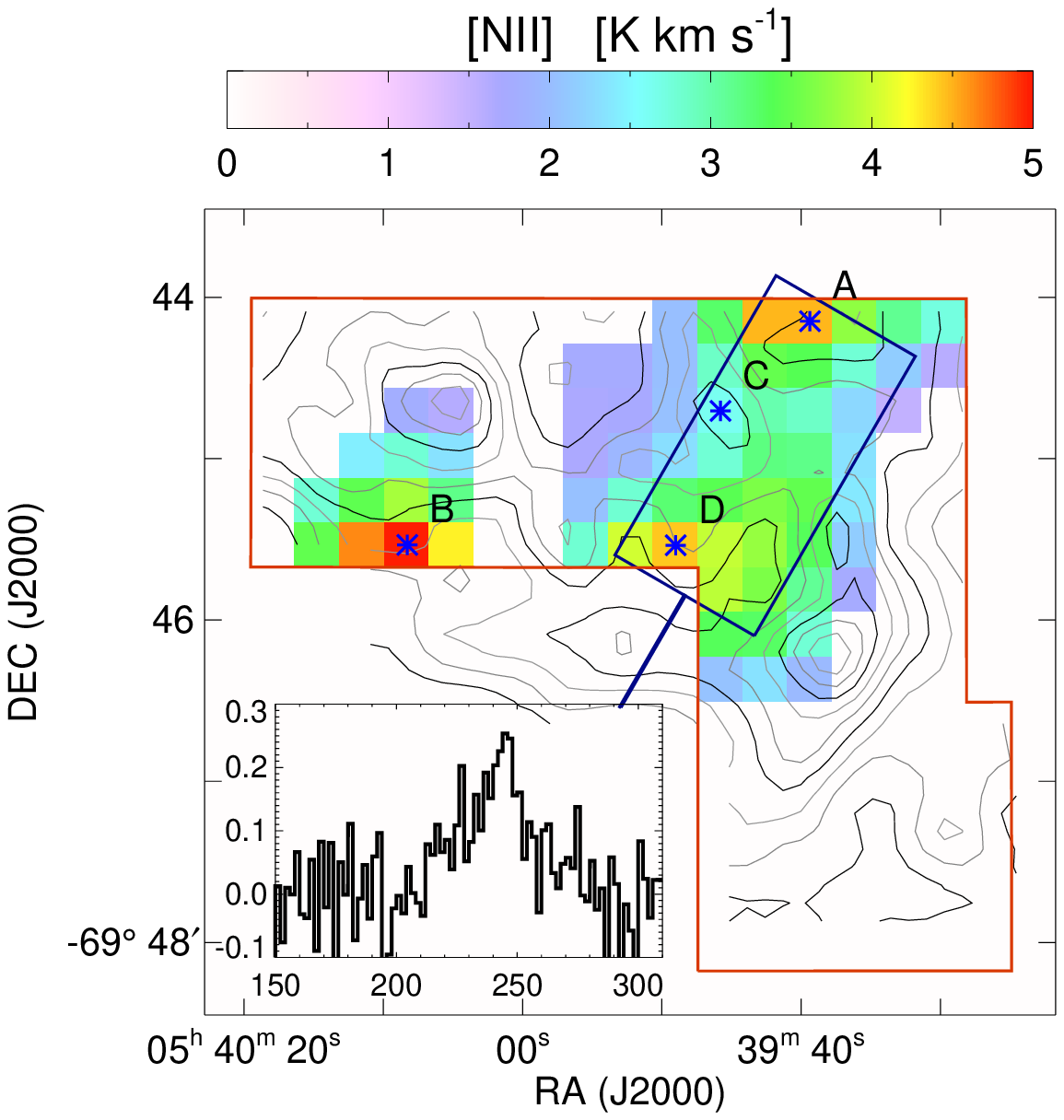}
\caption{Integrated (220 - 250~\kms) intensity map of \nii\ (colors), convolved to 50\arcsec\ angular resolution, overlayed with the contours of \cii\ at 20\arcsec\ resolution.  Blue asterisks show the positions where spectra have been extracted for Fig.~\ref{fig:NIICIIspectra}.  The inserted spectrum shows the \nii\ emission averaged in the area marked by the blue box ($x$-axis is Velocity [\kms] and $y$-axis is $T_\textrm{mb}$~[K]). \label{fig:NII_integmap}}
\end{figure}

We  examine the ionized gas contribution to the \cii\ emission using the \nii\ 205\um\ emission.  Since the \nii\ emission is very weak, we convolve the \nii\ emission to a spatial resolution of 50\arcsec\ and a spectral resolution of 2~\kms.  The integrated intensity map (Fig.~\ref{fig:NII_integmap}) shows an anti-correlation between the \cii\ and \nii\ emission, which indicates that most of the emission does not come from the ionized gas, which also emits in \nii.  To estimate the contribution of the ionized gas to the \cii\ emission quantitatively, we compare the observed \nii/\cii\ ratio with predictions based on a simple isothermal three-level system, for electron temperatures in the range 7000--12000~K.  In Table~\ref{table:niicii}, the observed ratios (in 50\arcsec\ beams) are listed for the four positions indicated in Fig.~\ref{fig:NII_integmap}, as well as the average over the blue box area in Fig.~\ref{fig:NII_integmap} and the average over the whole observed area.  For the calculations we used the elemental abundances of C and N for LMC \hii\ regions as given by \citet{Garnett1999}. The N abundance is consistent with \citet{Vermeij2002}.  We assume optically thin emission and LTE conditions (high density limit).  These assumptions provide an upper limit on the contribution of the ionized gas to the \cii\ emission for the following reasons.  When we calculate \nii\ 205\um/\cii\ 158\um\ as a function of the electron density, it has a maximum around 50--100~\cc\ and reaches its minimum (0.074) in the high density limit. The  \nii\ 205\um/\cii\ 158\um\ ratio is much less sensitive to the electron density than \nii\ 122\um/\cii\ 158\um; the former varies by only $\sim 65$\% over the whole electron density range, whereas the latter varies by one order of magnitude.  Still, if the electron density is at a range where the predicted \nii\ 205\um/\cii\ 158\um\ ratio is higher than 0.074, the \cii\ intensity attributed to the ionized gas becomes lower.  If the observed \cii\ emission is optically thick (or $\tau\sim 1$, see previous section), the optical-depth-corrected \cii\ emission would be stronger, so the fraction of the ionized gas contribution would decrease.  The maximum possible contribution from the ionized gas to the \cii\ emission is $40$--$60$\% at positions A, B, and D, where the \nii\ emission has a peak, with only 20\% at position C.  Position C is the \cii\ peak, which has only very weak corresponding CO and \ci\ emission (see p7 in Sect.~\ref{subsec:result_structure}).  Using the \nii\ emission, we exclude the ionized gas as a dominant origin of the \cii\ emission at this position.  The average contribution over the blue box in Fig.~\ref{fig:NII_integmap}, where the \nii\ is detected (at 50\arcsec\ resolution), is 30\%, and over the whole observed area it is 15\% (Table~\ref{table:niicii}).

\begin{figure}
\centering
\includegraphics[width=\hsize]{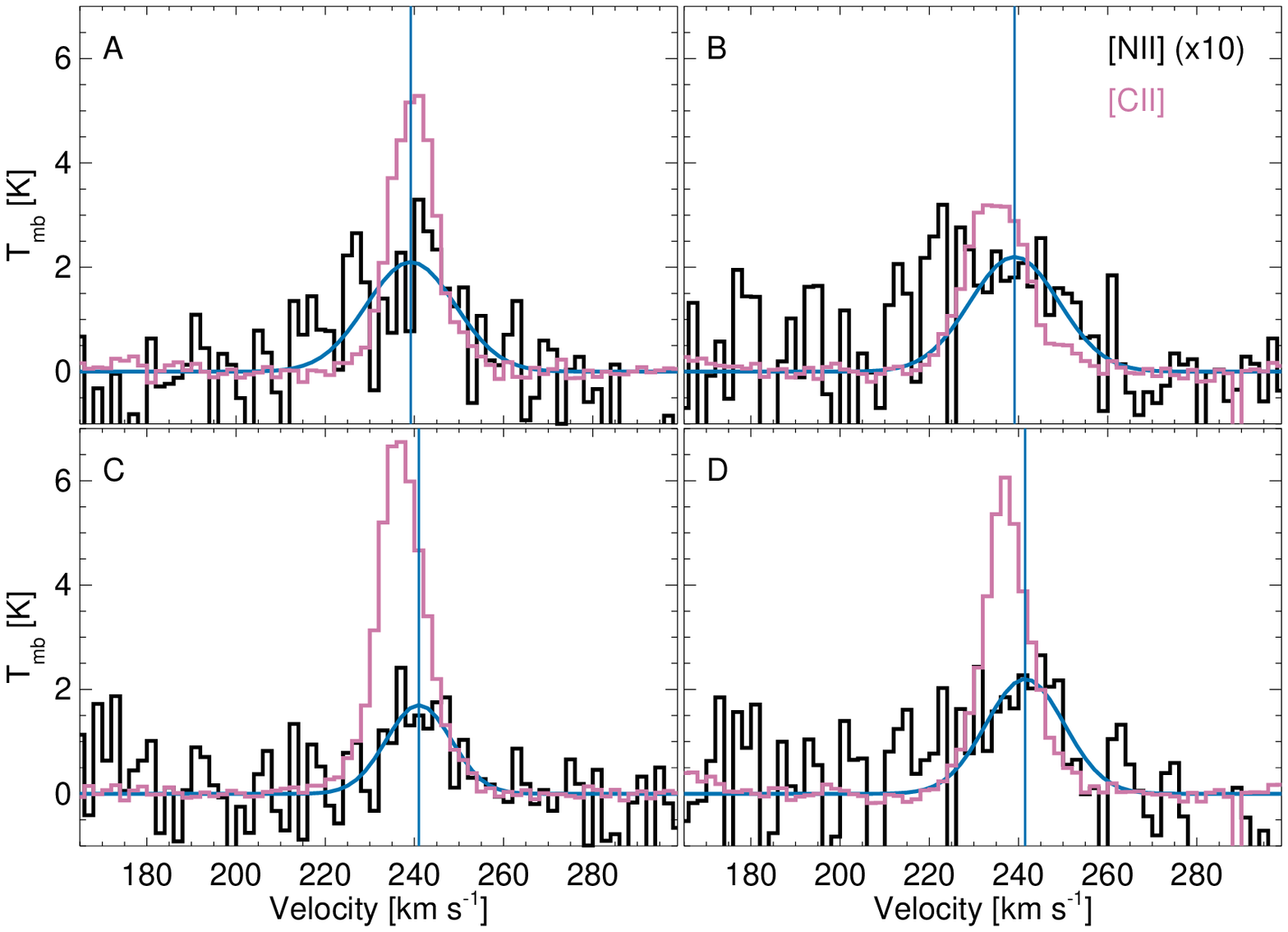}
\caption{\cii\ (purple) and \nii\ (black; scaled by 10) spectra at positions indicated in Fig.~\ref{fig:NII_integmap} with 2~\kms\ spectral resolution.  Blue lines show the result of a single Gaussian fit to the \nii\ emission, and its center is indicated as a vertical line.\label{fig:NIICIIspectra}}
\end{figure}

In Fig.~\ref{fig:NIICIIspectra} the velocity profiles of \cii\ and \nii\ at these four positions are compared.  We fit the \nii\ line with a single Gaussian, and the fitted central velocity is indicated as a vertical line.  For positions A and B, the difference of the velocity profile between the \nii\ and \cii\ is within the noise, but for positions C and D, \nii\ has a higher central velocity, and the \cii\ profile of the higher velocity wing matches the \nii\ profile, indicating that this velocity component might be associated with the ionized gas. The ratio of \nii/\cii\ in the high-velocity wing ($>245$~\kms) for position C and D is consistent with their origin in the ionized gas.

The \cplus\ column density in the ionized gas can be directly estimated by the H$^+$ column density ($N$(H$^+$)) derived from the radio continuum observations.  \citet{MartinHernandez2005} derived the emission measure and the electron density at three compact peak positions by the 6~cm continuum observations using the ATCA.  These are the compact \hii\ regions identified by \citet{Indebetouw2004} and match the \cii\ peaks (blue asterisks in the CO(4-3) panel of Fig.~\ref{fig:integ_map}; p4, p11, and northeast of p1).  The electron density ($n_e$) at these positions is ($2.2$--$3.0$)$\times 10^3$~\cc, and together with the emission measure it gives $N$(H$^+$) of ($3$--$9$)$\times 10^{21}$~cm$^{-2}$, and thus $N$(\cplus) of ($3$--$7$)$\times 10^{17}$~cm$^{-2}$ if we assume that all carbon atoms are singly ionized.  Since two of them are spatially resolved by ATCA having the source size of 4--5\arcsec\ and the other one is smaller than the beam size ($<2$\arcsec), the dilution factor in our 20\arcsec\ map is 16--103, which makes the estimate of $N$(\cplus) $=4\times 10^{16}$~cm$^{-2}$ for p11 and northeast of p1, and $3\times 10^{15}$~cm$^{-2}$ for p4.  They are 2\% or less compared to the $N$(\cplus) derived from the \cii\ emission (Table~\ref{table:NH}). However, this should be interpreted as a contribution from the dense ionized gas traced preferably by the radio continuum (its intensity is proportional to $n_e^2$), but not from the diffuse ionized gas. The H$\alpha$ map by the Magellanic Cloud Emission-Line Survey (MCELS) shows a different morphology including more extended structure and additional peak positions \citep{Chen2010}.  Our \nii\ map does not have a peak at these compact \hii\ regions, but correlates with H$\alpha$, confirming that they instead trace more diffuse ionized gas than the radio continuum because the critical density of the \nii\ 205\um\ is 180~\cc\ \citep{Shibai1992}.  Since the critical density of the \cii\ 158\um\ with the collision partner of electron is 40 \citep{Shibai1992}, it also traces more diffuse ionized gas rather than dense ionized gas.  Thus, the \nii\ 205\um\ emission is useful for estimating the ionized gas contribution to the \cii\ emission, tracing the same gas.  The reason that the \cii\ emission has peaks at the positions of compact \hii\ regions should be interpreted  as a dominant contribution of the PDR gas that surrounds these dense \hii\ regions and is illuminated by strong the UV radiation.

Theoretical models of \hii\ region/PDR complexes of solar metallicity gas predict contributions from the ionized gas to the \cii\ emission as high as 40\% \citep{Abel2005} or less than 10\% \citep{Kaufman2006} for most of the physical parameter range.  Observations toward Galactic PDRs have in fact provided comparable results.  Towards the Carina nebula, \citet{Oberst2011} derive 37\% from the \nii\ 205\um/\cii\ 158\um\ ratio, while using the correlation of \cii\ 158\um\ with \nii\ 122\um\ and \oi\ 63\um\ \citet{Mizutani2004} find 20\%.  \citet{Giannini2000} show a 10--30\% contribution from the ionized gas in NGC~2024 by CLOUDY modeling.  In S171 and $\sigma$ Sco region, the PDR contribution to the \cii\ emission is estimated by the PDR modeling, and the ionized gas contribution shows a spatial distribution with about 80\% at positions closer to the exciting stars, but only 20\% or 30\% at positions farther away \citep{Okada2006,Okada2003}.  Similarly 10-30\% contribution of the ionized gas is reported in several star-forming regions in nearby galaxies with almost solar metallicity \citep{Contursi2013,Kramer2005}.  \citet{Lebouteiller2012} present that 95\% of the \cii\ emission arises in PDRs in LMC-N11B, except toward the star cluster for which as much as 15\% could arise in the ionized gas.  In M33 star-forming regions a typical ionized gas contribution is estimated to be 5-50\% \citep{Higdon2003}.  In principle in a lower metallicity environment, the \cii\ emission from the \hii\ region decreases because the size of the \hii\ region is independent of the metallicity and the metal emission line is proportional to the abundance, while the column of \cplus\ layer in a PDR stays constant.  Therefore, the ionized gas contribution in a lower metallicity environment should be lower \citep{Israel2011,Kaufman2006}.  However, because of the uncertainties of the results cited above, this dependency has not been demonstrated by observations.

\begin{table}
\caption{Maximum possible contribution of the ionized gas to the \cii\ emission based on the integrated intensity ratio of \nii/\cii.\label{table:niicii}}
\centering
\begin{tabular}{ccc}
\hline\hline
Position & \nii/\cii & Max. ionized \\
& & gas fraction [\%] \\
\hline
A & $0.031\pm 0.004$ & $     44\pm    5$  \\
B & $0.043\pm 0.006$ & $     61 \pm   8$  \\
C & $0.014\pm  0.002$ & $     19\pm    3$  \\
D & $0.028\pm  0.005$ & $     39 \pm   6$  \\
\hline
(average) &&\\
\ \ blue box in Fig.~\ref{fig:NII_integmap} & $0.022\pm  0.002$ & $     30\pm    3$  \\
\ \ whole region & $0.011\pm  0.001$ & $     15 \pm   2$  \\
\hline
\end{tabular}
\end{table}

\subsection{Spatially resolved $I_\textrm{[CII]}$/$I_\textrm{FIR,th}$}

\begin{figure}
\centering
\includegraphics[bb=40 0 424 355,width=\hsize,clip]{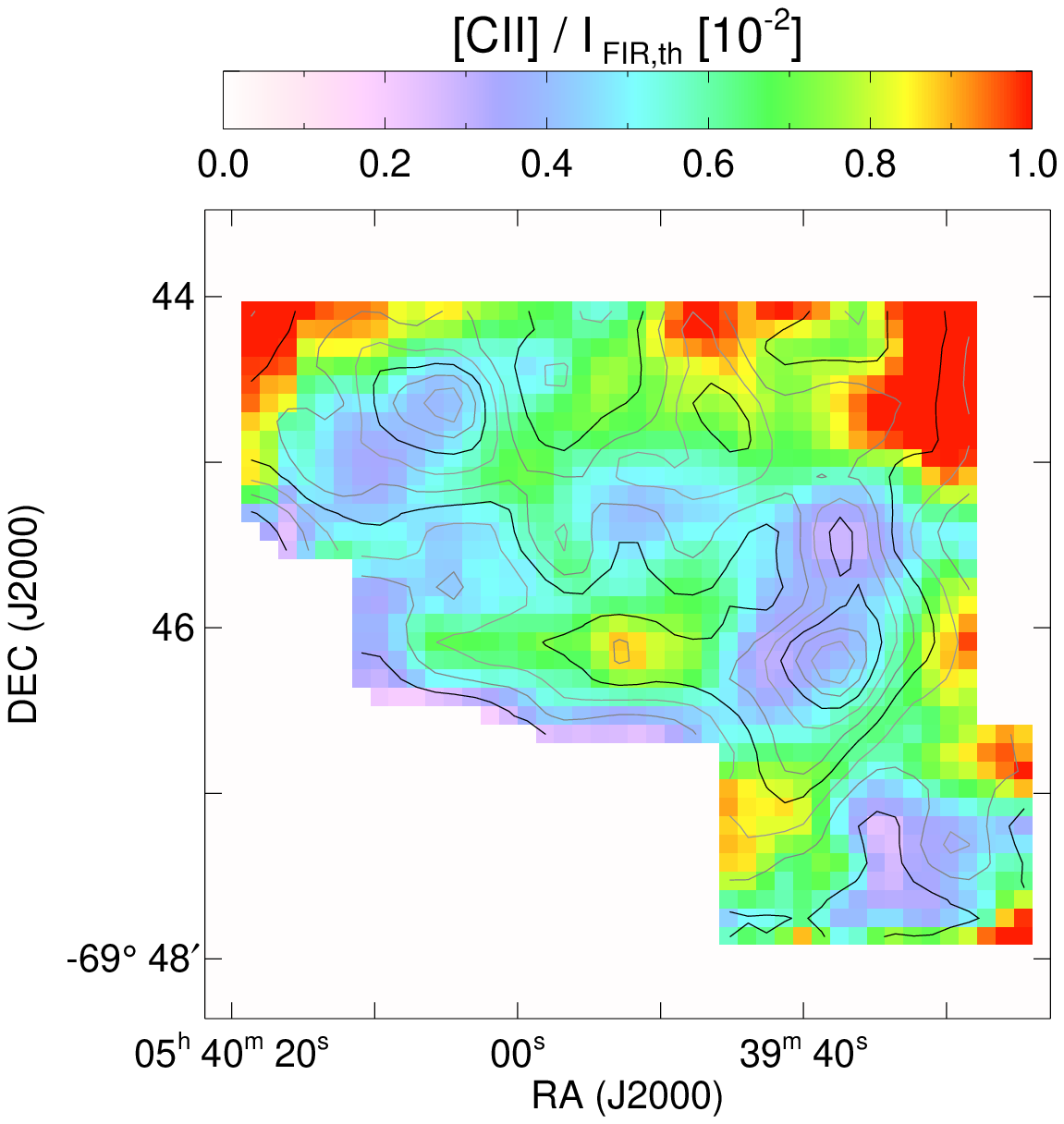}
\caption{\cii/$I_\textrm{FIR,th}$ map (see text for the definition) overlayed with the contours of the \cii\ integrated intensity.\label{fig:CIIdivFIR}}
\end{figure}

The ratio \cii/$I_\textrm{FIR}$ has been widely used as a tracer of the photoelectric heating efficiency, under the assumption that \cii\ 158\um\ is the dominant cooling line.  Here we derive the spatial distribution of this ratio.  We obtained the archival data of MIPS 24\um\ (the legacy survey of SAGE), PACS 100\um\ and 160\um, and SPIRE 250\um\ (the open key time program HERITAGE); applied the color correction; convolved them to a spatial resolution of 20\arcsec; fitted them with a modified blackbody with a single temperature and a spectral index of $\beta=1.5$; and integrated the fitted curve over the total range of the thermal emission.  Traditionally, $I_\textrm{FIR}$ is defined as the 42\um--122\um\ flux and $I_\textrm{TIR}$ is defined as the 3\um--1100\um\ flux \citep{DaleHelou2002}.  Here we use the direct integration of the fitted total thermal emission, and express it as $I_\textrm{FIR,th}$.  We do not calculate $I_\textrm{TIR}$ because we do not have a spectrum of polycyclic aromatic hydrocarbons (PAHs) at each position, and its contribution to total infrared flux (TIR) can vary in spatially resolved regions.  We do not adopt $I_\textrm{FIR}$ either because there is no reason to cut the integration at 122\um.  Figure~\ref{fig:CIIdivFIR} shows the spatial distribution of $I_\textrm{[CII]}$/$I_\textrm{FIR,th}$.  The ratio is significantly lower at the N159~W and E cores, being $0.3$--$0.4$\%.  At the \cii\ peak between N159~W and E (p7), $I_\textrm{[CII]}$/$I_\textrm{FIR,th}=0.7$--0.8\%.  These values are in the same range as the results in the LMC/N11 region by \citet{Lebouteiller2012} when considering that TIR, which includes mid-infrared emissions, is 10--35\% higher than the total thermal emission \citep[estimate among the sample of PDRs in][ except for Ced~201, which is excited by a much later star of B9.5]{Okada2013}.  \citet{Israel1996} present a value of $I_\textrm{[CII]}$/$I_\textrm{FIR}$ of 1--2\% for the  N159~E and W cores.  Their much larger value is due to the definition of $I_\textrm{FIR}$ and the lower spatial resolution (the FIR emission has a sharper peak at the cores than \cii).

There are three possible causes of the lower $I_\textrm{[CII]}$/$I_\textrm{FIR,th}$ at the cores.  The first possibility is a contribution of the \oi\ 63\um\ to the cooling.  Here we do not directly use the archival PACS spectroscopic data because they do not cover our whole observed region, but the comparison of the \oi\ 63\um\ and \cii\ 158\um\ emissions within their observed area indicates that the intensity of \oi\ 63\um\ is 30--100\% of the \cii\ 158\um\ intensity.  We plan to observe velocity resolved \oi\ 63\um\ with GREAT at our next southern flight opportunity and  carefully compare the line profile.  The second possibility is that the photoelectric heating efficiency is actually low at these cores.  The photoelectric heating efficiency is a function of the charging parameter $\gamma=G_0 T^{1/2}/n_e$, where $G_0$ is UV radiation field, $T$ is the gas temperature, and $n_e$ is the electron density \citep{Bakes1994}.  When $\gamma$ is high, dust grains are positively charged.  Thus, the photoelectric heating efficiency becomes lower because a higher energy is needed to further ionize them.  If there is a strong UV source in the core, $\gamma$ in the core is high and the heating efficiency could be lower than outside.  This may be likely because both N159~E and W possess compact \hii\ regions (Fig.~\ref{fig:integ_map} CO(4-3) panel).  Correlation with the fraction of positively ionized PAHs would give more detailed examination of this possibility \citep{Okada2013}, but there are only sparse Spitzer/IRS observations in this region.  The third possibility is that later-type heating sources contribute to $I_\textrm{FIR,th}$ but not to the photoelectric heating because of the lack of photons with the energy of $>6$~eV.

\section{Summary}

We mapped a $4\arcmin\times (3\arcmin$--$4\arcmin)$ region in the N159 star-forming region in the LMC in the \cii\ 158\um\ and \nii\ 205\um\ lines with the GREAT instrument on board SOFIA, as well as CO(3-2), (4-3), (6-5), \coiso(3-2), and \ci\ \transl\ and \transu\ with  APEX, and study their velocity profiles in detail.  The variation of the velocity structure over the observed area is large, and the \cii\ emission shows a significantly different velocity profile from the other lines at most positions.  Quantitatively, we estimate the fraction of the \cii\ emission that cannot be attributed to the material traced by the CO line profiles to be 20\% around the CO cores and up to 50\% in the area between the cores.  We derived the column density of \cplus, C, and CO in each velocity bin, and find that the relative contribution from \cplus\ dominates at the velocity range far from the line center, and the area between the CO cores.  The origin for this \cii\ emission, at velocities substantially off the velocities of the dense molecular cores traced by CO and \ci, may be due to gas ablating from the dense cloud cores or  to additional gas not immediately associated with the dense cores.

To estimate the contribution of the ionized gas to the \cii\ emission, we use the ratio with the \nii\ 205\um\ line and show that the ionized gas can contribute up to 19\% to the \cii\ emission at its peak position, less than 30\% in the area where \nii\ is detected, and less than 15\% if averaged over the whole area observed.  By comparing the peak velocity of the \cii\ and \nii\ emissions, we provide evidence that a high-velocity wing in the \cii\ emission is likely due to ionized gas.  Our study emphasizes that it is essential to resolve the velocity profile in order to distinguish between different cloud components and correctly derive the physical properties of individual components.

%-------------------------------------------------------------------

\begin{acknowledgements}
This work is based in part on observations made with the NASA/DLR Stratospheric Observatory for Infrared Astronomy (SOFIA). SOFIA is jointly operated by the Universities Space Research Association, Inc. (USRA), under NASA contract NAS2-97001, and the Deutsches SOFIA Institut (DSI) under DLR contract 50 OK 0901 to the University of Stuttgart. This work is carried out within the Collaborative Research Centre 956, sub-project A3, funded by the Deutsche Forschungsgemeinschaft (DFG).  We thank F. Israel for contributing to the FIFI-GREAT comparison.  We thank C.-H.R. Chen for providing a list of the OB star candidates.

\end{acknowledgements}

\end{document}